\documentclass[fleqn,usenatbib]{mnras}

\usepackage{newtxtext,newtxmath}
\usepackage{lastpage}
\usepackage{ulem}
\usepackage[T1]{fontenc}

\DeclareRobustCommand{\VAN}[3]{#2}
\let\VANthebibliography\thebibliography
\def\thebibliography{\DeclareRobustCommand{\VAN}[3]{##3}\VANthebibliography}
                                                                   
\usepackage{graphicx}	
\usepackage{amsmath}	

\title[Probing the HI distribution using 21-cm IM]
{Probing the HI distribution at small scales using 21-cm Intensity Mapping at large scales}

\author[M. Chhabra and S. Bharadwaj]{
Minal Chhabra$^{1}$\thanks{E-mail: minal99chhabra@gmail.com} and Somnath Bharadwaj$^{1}$\thanks{E-mail: somnath@phy.iitkgp.ac.in}
\\
$^{1}$Department of Physics, Indian Institute of Technology Kharagpur, West Bengal 721302, India
}
\date{Accepted XXX. Received YYY; in original form ZZZ}

\pubyear{2026}

\begin{document}
\label{firstpage}
\pagerange{\pageref{firstpage}--\pageref{lastpage}}
\maketitle

\begin{abstract}
Neutral hydrogen (HI)  21-cm Intensity Mapping (IM) holds the potential to map the large-scale structures in the Universe over a wide redshift range $(z \lesssim 5.5)$, measure cosmological parameters, and shed light on the nature of dark energy. In addition, the signal is also sensitive to how the HI is distributed among the dark matter haloes, this being quantified through the HIHM relation, which relates the HI mass to the halo mass.  In this work, we investigate whether measurements of the 21-cm power spectrum (PS) and bispectrum (BS) at large scales can be used to estimate the HIHM relation, which quantifies the HI distribution at small scales. As a proof of concept, we consider the simulated 21-cm IM signal at $z=1$. We find that the measured 21-cm PS and BS at large scales $(k \le k_{ul} = 0.32 \, {\rm Mpc}^{-1})$ are well modeled using perturbation theory, with only two free parameters namely $[\Omega_{\rm HI} b_1]$ and $\gamma = b_2/b_1$.  Combining the measured 21-cm PS and BS with an independent measurement of $\Omega_{\rm HI} $, we show that it is possible to estimate the three parameters that quantify the HIHM relation.  We expect observational estimates of the HIHM relation to shed light on galaxy formation and the evolution of the ISM. 
Our preliminary analysis ignores redshift space distortion and the system noise in IM observations, which we plan to address in future work. 
\end{abstract}


\section{Introduction}
\label{sec:intro}

Intensity Mapping (IM) with the neutral hydrogen (HI) 21-cm line \citep{bharadwaj2001using, bharadwaj2001hi} holds great promise in observing the universe across a wide redshift range \citep{bharadwaj2005using, pritchard201221}.  In the post-reionization (Post-EoR) era $(z \lesssim 5.5)$, the HI is mostly associated with galaxies that reside in dark matter haloes, and we expect the 21-cm IM signal to trace the large-scale structures in the universe. Observations of the 21-cm IM signal are expected to shed light on the nature of dark matter and dark energy through BAO measurements \citep{chang2008baryon}. These observations also hold the potential to constrain cosmological parameters \citep*{bharadwaj2009estimation} even without measurements of the BAO. Furthermore, 21-cm observations also have the potential to quantify non-Gaussianity, both induced  \citep{bharadwaj2005using} and primordial \citep{hazra2012primordial}. 

Several observational efforts have been made to measure the Post-EoR 21-cm IM signal. There have been several detections of the cross-correlation power spectrum (PS) of the 21-cm IM signal and galaxy surveys at $z \le 1$ \citep{chang2010intensity, masui2013measurement, switzer2013determination, anderson2018low, wolz2016intensity,tramonte2020neutral, cunnington2023h, amiri2023detection, amiri2024detection}. The first attempts to measure  the 21-cm IM PS (auto-correlation) were made by \cite{ghosh2011gmrt, ghosh2011improved} at $z=1.32$ using the Giant Metrewave Radio Telescope (GMRT, \cite{swarup1991giant}).  A series of works  \citep{chakraborty2021first,pal2022towards,elahi2023towards,elahi2023towards1, elahi2024towards} have attempted to measure the 21-cm IM PS in the redshift range $1.9 \lesssim z \lesssim 2.6$ using data from the upgraded GMRT\footnote{\url{https://www.gmrt.ncra.tifr.res.in/}} \citep{gupta2017upgraded}, which led to an upper limit of $[\Omega_{\rm HI} b_{\rm HI}] < 0.011$.
\citet{paul2023first} have recently reported the first detection of the  21 cm PS at redshifts $z = 0.32$ and $0.44$ using the MeerKAT\footnote{\url{https://www.sarao.ac.za/science/meerkat/about-meerkat/}} interferometer  \citep{gibbon2015fiber}. \citet{mazumder2025hi} have reported upper limits on the 21-cm PS at redshifts $0.4 < z < 0.48$  from the MeerKAT International GigaHertz Tiered Extragalactic Exploration (MIGHTEE) survey. 
{\cite{chime-auto-paper} have reported a $12 \sigma$ detection of the 21-cm PS at $z=1.16$ using Canadian Hydrogen Intensity Mapping Experiment (CHIME\footnote{\url{https://chime-experiment.ca/en}}, \cite{bandura2014canadian}). A recent low-redshift detection $(0.02 \lesssim z \lesssim 0.07)$ using the data from MIGHTEE survey has also been repoted by \cite{townsend2026measurements}.}
Several existing and upcoming telescopes, including the 
Hydrogen Intensity and Real-time Analysis eXperiments (HIRAX\footnote{\url{https://hirax.ukzn.ac.za/}}, \cite{newburgh2016hirax}), and the Square Kilometer Array (SKA)-MID\footnote{\url{https://www.skao.int/en/explore/telescopes/ska-mid}}, aim to measure the Post-EoR 21-cm IM signal with great sensitivity.

There is considerable interest in modeling the Post-EoR HI distribution. Such models are necessary to predict the expected 21-cm IM signal and also to interpret the observations once the signal has been detected. Several attempts have been made to analytically model Post-EoR HI distribution \citep{marin2010modeling, padmanabhan2016constraining, padmanabhan2017halo, penin2018scale}. There have also been works that numerically simulate the 21-cm IM signal using N-body simulations \citep{bharadwaj2004hi},  and also hydrodynamical simulations \citep{dave2013neutral, villaescusa2014modeling, villaescusa2018ingredients}. In this work, we primarily focus on a semi-numerical scheme that starts from a simulated distribution of dark matter haloes, and simulates the 21-cm IM signal by populating the haloes with HI using an HI mass -- Halo mass (HIHM) relation, which relates the HI mass to the halo mass. Several studies have used such semi-numerical simulations to address various aspects of the expected Post-EoR 21-cm IM signal \citep{bagla2010h, khandai2011detecting, guha2012constraining, sarkar2016modelling, seehars2016simulating, sarkar2018modelling, sarkar2019redshift, modi1904intensity, wang2021breakdown}. Here we particularly note \cite{sarkar2019modelling} who have simulated the 21-cm PS and bispectrum (BS) over a wide redshift range $(1 \le z \le 6)$, and used this to study the $z$ evolution of the linear and quadratic bias parameters. We also mention several other works that have modeled the 21-cm BS \citep{saiyad2006probing, durrer2020full, cunnington2021h, raste202421}.

As mentioned earlier, in the Post-EoR era,  the HI is mostly associated with galaxies that reside within dark matter haloes. The HIHM relation quantifies how the HI is distributed among the haloes. This is of considerable interest, not only for predicting the expected 21-cm IM signal,  but also as a diagnostic for galaxy formation and the evolution of the interstellar medium (ISM). There have been several efforts to observationally study how the HI is distributed among the haloes, and also among the galaxies. Considering HI-selected galaxies, \cite{dutta2022dark} have utilized stacking to determine the HI abundance in haloes of varying masses, whereas  \citet{guo2017constraining} and \citet{paul2018halo} have applied halo occupation distribution (HOD)-like prescriptions to model the HI distribution. Note that the 21-cm emission from individual galaxies is too faint to be detected by existing radio telescopes for $z \ge 0.1$, unless the galaxy is gravitationally lensed \citep{saini2001using, chakraborty2023detection}. It is necessary to consider Lyman-$\alpha$ absorption studies (e.g. \cite{rhee2018neutral}) to directly access the HI content at $z \ge 0.1$. Several studies have employed analytical halo model prescriptions to analyze the HI content in galaxies and damped Lyman-$\alpha$ systems (DLAs) to determine the HIHM relation  \citep{padmanabhan2016constraining, wolz2016intensity, padmanabhan2017halo, padmanabhan2017constraints}.  In a recent work, \cite{padmanabhan2023hi} has studied the implications of the recently claimed 21 cm PS detection \citep{paul2023first} on the redshift evolution of the HI distribution. In summary, here, we note that there is considerable interest, motivation, and scope for studying the HIHM relation across the entire redshift range $z < 6$. Such studies will not only be useful for utilizing the Post-EoR 21-cm IM signal to estimate cosmological parameters, but they will also shed light on various galaxy formation and ISM processes. 

In this work, we study whether measurements of the 21-cm PS and BS at large length scales can be used to probe the HIHM relation, which tells us how the HI is distributed among haloes of different masses. As a proof of concept, we demonstrate this at $z=1$ using a simulated 21-cm IM signal. The preliminary analysis presented here ignored redshift-space distortions and system noise in IM observations, which we plan to address in future work. 
We have simulated the expected 21-cm signal using the HIHM relation proposed by \cite*{padmanabhan2017halo}, and we investigate whether it is possible to estimate the parameters of this HIHM relation by modeling the 21-cm PS and BS. Here, we employ the two complimentary techniques to model the 21-cm PS, namely, the standard perturbation theory and the halo model formalism. These techniques differ in how they estimate the HI density parameter and the linear bias: one relies on a simulation-based approach, whereas the other uses an analytical halo-model calculation. The 21-cm BS, however, has been modeled perturbatively throughout.
A brief outline of the paper follows. 

In Section~\ref{sec:HI model}, we describe how we have modeled the HI distribution using the HIHM relation, and used it to simulate the expected $z=1$ 21-cm IM signal. In Section~\ref{sec:Bias est} we quantify the statistics of the 21-cm IM signal using the PS and BS. We model these using perturbation theory and a bias prescription, and present estimates for the bias parameters.  We present and demonstrate our method to estimate the parameters of the  HIHM relation in Section~\ref{sec:param est}, and consider the halo model approach in Section~\ref{subsec:halomod}.  We summarize and discuss our work in Section~\ref{sec:summary}.

In this work, we have used the $\Lambda$CDM cosmological model with parameters taken from \cite{refId0}. Also, we have used the transfer function given by \cite{eisenstein1999power} to generate the linear matter PS. 

\section{Modeling the HI Distribution}
\label{sec:HI model}

The absence of Lyman-$\alpha$ absorption troughs in low redshift quasar (QSO) spectra \citep{GP1965,Scheuer1965} indicates that the diffused IGM is completely ionized at redshifts $z \lesssim 5.3$ (e.g. \citealt{zhu2024damping}).
It is believed that, at these redshifts, the HI primarily resides in galaxies, where it is self-shielded against the radiation that ionizes the IGM. At low redshifts $(z \lesssim 0.1)$, HI is directly observed in galaxies using 21 cm radio emission (e.g. \citealt{zwaan2005hipass, martin2010arecibo}). At higher redshifts ($z \sim 2-5$), most of the HI is thought to lie in damped Lyman-$\alpha$ systems (DLAs) observed in the absorption spectra of high redshift QSOs (see \cite{rao2005damped} for a review). The HI column densities of these DLAs ($N_\text{HI} > 10^{20} \mbox{ cm}^{-2}$) are comparable to that expected of galaxies  \citep{lanzetta1991new}. DLA observations indicate $\Omega_{\rm HI} $  the HI density parameter to have a value $\Omega_{\rm HI} \approx (0.4 - 1) \times 10^{-3}$ in the redshift range $z \sim 1.9 - 2.6$ \citep{bird2017statistical, rhee2018neutral}. Using the HI galaxy stacking technique, \cite{chowdhury2020h} have found $\Omega_{\rm HI} \approx 0.45 \times 10^{-3}$ at an average redshift of $z=1.06$. Constraints on $\Omega_{\rm HI}$ are also obtained from HI galaxy surveys and intensity mapping using the cross-correlation with optical surveys at very low redshifts $z \lesssim 0.2$ (see \cite{padmanabhan2015theoretical} for a compilation).

It is useful to quantify the HI distribution using the HI mass -- halo mass relation (HIHM) that relates  $M_\text{HI}$ the total HI mass contained in a galaxy to $M_h$ the mass of the dark matter halo that hosts the galaxy.  In an early work, \cite*{bagla2010h} proposed an HIHM relation that assumes $M_\text{HI}$ to be proportional to $M_h$ within a range of halo masses $M_{\rm min} \le M_h \le M_{\rm max}$. The redshift-dependent halo mass cutoffs correspond, respectively, to fixed circular velocity limits of $30$ and $200 \, {\rm km \, s^{-1}}$  determined using DLA simulations \citep{pontzen2008damped}. \cite{bagla2010h} also propose two other variants of the HIHM relation that replace the abrupt cut-off with a smooth cut-off at the high mass end. 

In the present analysis, we adopt a more generalized HIHM relation proposed by  \citet*{padmanabhan2017halo}. This has three parameters, namely $\alpha$, $\beta$, and $v_{c0}$, which correspond respectively to the normalization, excess logarithmic slope, and lower cutoff on circular velocity. It is convenient to write the HIHM relation using   
 \begin{equation}
    M_\text{HI} = \alpha f_{H,c} M_h \left( \frac{M_h}{10^{11} h^{-1} M_\odot} \right)^\beta \exp \left[-\frac{M_\text{cut}}{M_h}\right]
    \label{eq:HImodel}
\end{equation} 
where 
\begin{equation}
    M_\text{cut} = 10^{10} M_\odot \left[\frac{v_{c0}}{60 \, {\rm km \, s^{-1}}} \right]^3   \left(\frac{1+z}{4} \right)^{-\frac{3}{2}} 
\end{equation}
and 
$f_\text{H,c} = (3/4) \Omega_b/\Omega_m$. The model also includes an HI density profile inside the halo. However, our analysis is restricted to large length scales, where this can be ignored. 
In an earlier work, \citet{padmanabhan2016constraining}  also included an upper mass cutoff for $M_h$ to account for the fact that the more massive haloes that host large elliptical galaxies and clusters do not contain much HI \citep{serra2012atlas3d}. However, it was found that the preferred value of the upper mass cutoff is predicted to be very large, so much so that it makes an insignificant contribution to the large-scale HI distribution and may be ignored. 

\cite{padmanabhan2017halo} have obtained constraints on the parameters using observations of HI mass function and intensity mapping results at low redshifts ($z = 0-1$) and DLA incidence and column density distributions at high redshifts ($z=2.3-5$). The best-fit values obtained for the parameters are $\alpha = 0.09$, $\beta = -0.58$, and $v_{c0} = 36.3 \, {\rm km \, s^{-1}}$, which translates to a mass cutoff $M_{\rm cut} = 6.26 \times 10^{9} M_\odot$ at $z=1$. We have used these best-fit parameters as fiducial values in our analysis. 

We have used the above HIHM relation to populate HI in the haloes obtained from a set of dark matter only cosmological simulations. The simulations are the same as \citet*{sarkar2016modelling}, except that we have augmented the previous five realizations with {$95$} additional realizations of the simulations, {resulting in a total of $100$ realizations that we have used for our analysis.}. Furthermore, the entire analysis presented here is restricted to a single redshift $z=1$. The simulations have a comoving volume of $[150.08 \mbox{ Mpc}]^3$ with a spatial resolution of $0.07 \mbox{ Mpc}$. The haloes are identified using the Friends-of-friends (FoF) algorithm, which sets the halo mass resolution to approximately $10^9 M_\odot$. 
{We place the entire HI assigned to each halo (eq. \ref{eq:HImodel}) at the position of its center and do not consider a density profile within the halo. This is because our analysis is restricted to large scales, where the effect of the HI profile inside halos is negligible. We have gridded the simulated HI distribution using the mass-weighted Cloud-in-Cell (CIC) algorithm \citep{hockney2021computer} to a spatial resolution of $\sim 0.56 \, {\rm Mpc}^{-1}$. The gridding does not affect the statistics significantly at large scales. We use the gridded HI field}
to obtain $ \rho_\text{HI}(\mathbf{x})$ the comoving HI density, which we have used to calculate the redshifted  21 cm brightness temperature using \citep{bharadwaj2005using},
\begin{equation}
    \label{eq:brit temp}
    \delta T_b (\mathbf{x},z) = \bar{T}(z) \mbox{ } \frac{\rho_\text{HI} (\mathbf{x})}{\bar{\rho}_\text{H}}
\end{equation}
where $\bar{\rho}_\text{H} = (3/4) \Omega_b \rho_{c0}$ is the mean hydrogen density  and  
\begin{equation}
    \bar{T} (z) = 4.0 \mbox{ mK } (1+z)^2 \left( \frac{\Omega_b h^2}{0.02} \right) \left( \frac{0.7}{h} \right) \frac{H_0}{H(z)} \,.
    \label{eq:T_bar}
\end{equation}
Note that the present analysis does not include redshift space distortion (RSD) due to peculiar velocities \citep{bharadwaj2001using,bharadwaj2004cosmic}, and is entirely restricted to "real space" as against "redshift space". We plan to consider the effect of RSD in future work.  

In the present work, we have considered the power spectrum (PS) and the bispectrum (BS) to respectively quantify the two-point and three-point statistics of the redshifted  21 cm brightness temperature signal. We have used ten statistically independent realizations of the simulated signal to estimate the mean and standard deviation for the PS and BS.

\section{Statistics and Bias Estimation}
\label{sec:Bias est}

\subsection{Power spectrum and bispectrum}
\label{subsec:PS BS}

\begin{figure}
\centering
    \includegraphics[width=0.9\columnwidth]{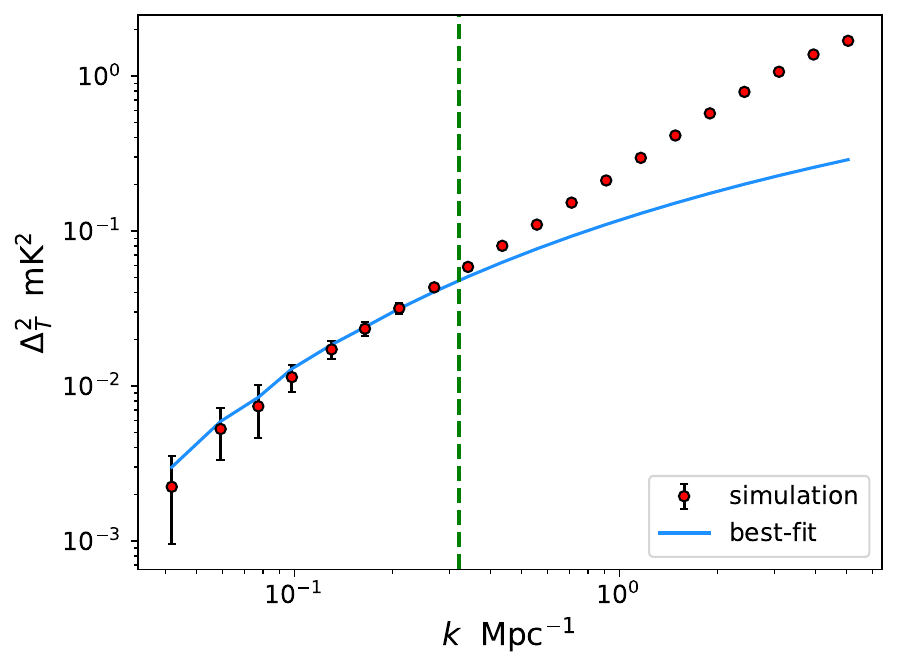}
    \caption{Mean-squared 21-cm brightness temperature fluctuations $\Delta_\text{T}^2 = k^3 P_\text{T}/2 \pi^2$ at $z=1$ for the fiducial values of HIHM model. Red points show the values obtained from the simulation, and error bars indicate the scatter in $100$ independent realizations. The solid line shows the predicted PS (eq. \ref{eq:ps}) calculated using the best-fit $[\Omega_{\rm HI} b_1]$, and the green dashed line indicates the extent of our fit ($k_{ul} = 0.32 \, {\rm Mpc}^{-1}$).}
    \label{fig:ps}
\end{figure}

We first consider $P_\text{T}(k)$ the PS of redshifted 21-cm brightness temperature fluctuations. Fig.~\ref{fig:ps} shows $\Delta_\text{T}^2(k) = k^3 P_\text{T}/2\pi^2$ the mean-squared brightness temperature fluctuations as a function of the comoving wave number $k$ in the range $0.04 \le k \le  5.05 \mbox{ Mpc}^{-1}$. The red points with error bars respectively indicate the mean and the standard deviations obtained from the simulated HI distribution. We see that  $\Delta_\text{T}^2(k) \approx 10^{-2} \, {\rm mK}^2$ at $k \approx 0.1  \mbox{ Mpc}^{-1}$ and it increases to  $\Delta_\text{T}^2(k) \approx 1 \, {\rm mK}^2$ at $k \approx 3  \mbox{ Mpc}^{-1}$, which indicates the level of the 21-cm brightness temperature fluctuations that we expect at $z=1$.   We note that the $1 \sigma$ error bars are extremely small at all $k$, except the smallest $k$ ($\le 0.1 \mbox{ Mpc}^{-1}$), given the volume of our HI simulations.

\begin{figure}
\centering
    \includegraphics[width=0.9\columnwidth]{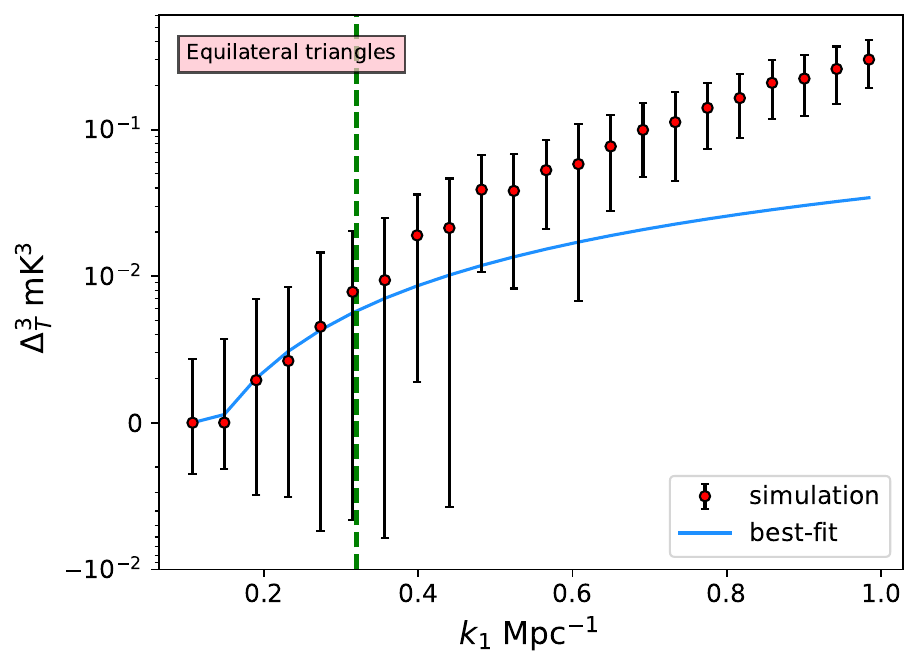}
    \includegraphics[width=0.9\columnwidth]{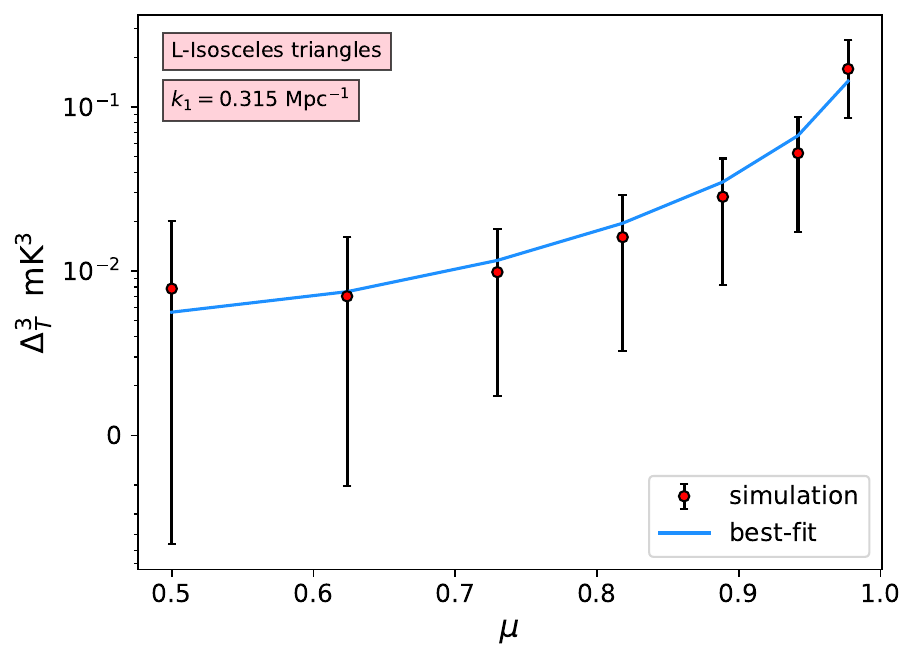}
    \caption{Mean-cubed 21-cm brightness temperature fluctuations $\Delta_\text{T}^3 = k^6 B_\text{T}/(2 \pi^2)^2$ at $z=1$ as a function of $k_1$ for equilateral triangles ({\it top}) and $\mu$ for L-isosceles triangles with $k_1 = k_2 = 0.315 \mbox{ Mpc}^{-1}$ ({\it bottom}). The red points are obtained from the simulation with error bars calculated using $100$ independent realizations. The solid line is the predicted BS (eq. \ref{eq:bs}) calculated using best-fit $[\Omega_{\rm HI} b_1]$ and $\gamma$. The green dashed line in the top panel indicates $k_{ul} = 0.32 \, {\rm Mpc}^{-1}$.}
    \label{fig:bs}
\end{figure}

We now consider $B_\text{T} (\mathbf{k_1},\mathbf{k_2},\mathbf{k_3})$ the BS of the redshifted 21 cm brightness temperature fluctuations, which is a function of three different k-modes that form a closed triangle ($\mathbf{k_1} + \mathbf{k_2} + \mathbf{k_3} = 0$). Note that the sides have been labeled such that $k_1\ge k_2 \ge k_3$. The value of the BS depends only on the shape and size of the triangle $(k_1,k_2,k_3)$, independent of how it is oriented. Here we parameterize the BS using  $B_\text{T} (k_1,\mu,t)$  \citep*{bharadwaj2020quantifying}, where $k_1$, the largest side of the triangle, parameterizes its size, and the dimensionless quantities  $\mu = -\mathbf{k_1} \cdot \mathbf{k_2}/ k_1 k_2$ and $t = k_2/k_1$  parameterize its shape. The parameters $(\mu,t)$ are restricted to the range  $ 0.5 \le (\mu,t) \le 1$ and the allowed region satisfies the constraint $2 \mu t \ge 1$ \citep{bharadwaj2020quantifying}. 
Fig.~\ref{fig:bs} shows the mean-cubed brightness temperature fluctuations, $\Delta_\text{T}^3 = k^6 B_\text{T}/(2\pi^2)^2$. The top panel shows $\Delta_\text{T}^3$  as a function of $k_1$ for equilateral triangles where  ${k_1} = {k_2} = {k_3}$ and $(\mu,t) \approx (0.5,1)$. The values of $k_1$ span the range $0.11 - 0.98 \mbox{ Mpc}^{-1}$, the red points show the values obtained from the HI simulation and the error bars are calculated using {$100$} independent realizations of the HI distribution. We see that $\Delta_\text{T}^3$ is consistent with zero for $k \leq 0.4 \mbox{ Mpc}^{-1}$ and increases to about $0.3 \mbox{ mK}^3$ for $k \approx 1 \mbox{ Mpc}^{-1}$. The bottom panel shows $\Delta_\text{T}^3$   as a function of $\mu$  for L-isosceles triangles $(t \approx 1)$  where the two largest sides are equal  ($k_1 = k_2 = 0.315 \mbox{ Mpc}^{-1}$). Here, $\mu=0.5$ corresponds to an equilateral triangle, the shape of the triangle changes as $\mu $ increases and, it approaches the squeezed limit $k_3 \rightarrow 0$ as $\mu \rightarrow 1$. We see that the value of the BS increases as we go from an equilateral to a squeezed triangle. Similar behavior is also seen for other values of $k_1$.  Here we have evaluated the full shape $(\mu,t)$ and size $k_1$ dependence of the BS $B_\text{T} (k_1,\mu,t)$. The full BS, for all possible triangle configurations, is shown in Fig.~\ref{fig:bs_all} of Appendix \ref{ap:full bisp}, where the shape and size dependence are discussed in more detail. 

\subsection{Bias prescription}
\label{subsec:bias-model}
On large scales, we may expect HI to trace the distribution of matter, albeit with a bias 
\citep*{bharadwaj2001using}. Here we adopt a perturbative bias prescription to relate  $\delta_\text{HI}$ the HI density contrast to $ \delta$ the matter density contrast. We then have {
\begin{equation}
    \delta_\text{HI} (\mathbf{x}) = b_1 \delta (\mathbf{x}) + \frac{b_2}{2} \delta^2 (\mathbf{x})
    \label{eq:bias}
\end{equation}
}
where $b_1$ and $b_2$ respectively are the linear and quadratic HI bias parameters. 
Similar bias prescriptions have been used extensively to model the large-scale galaxy distribution, and the reader is referred to \cite{desjacques2018large} for a review. 

{
Such a bias prescription along with Standard perturbation theory (SPT, \cite{goroff1986coupling,jain1993second, bernardeau2002large,carlson2009critical}) can be used to make analytical predictions for the PS and BS of the redshifted 21-cm brightness temperature fluctuations. Here, we use the linear model for the PS and the tree-level model for BS, which are the leading-order predictions. Note that for this case, the PS is independent of the quadratic bias. We have attempted to include higher-order (1-loop) terms that involve mode coupling and contributions from the quadratic ($b_2$) and cubic ($b_3$) bias parameters. However, we find that although such a prescription can reliably model the matter and halo distribution, it fails to simultaneously model the 21-cm PS and BS (see Appendix \ref{ap:modPSBS}). The fact that linear perturbation theory is preferred over the 1-loop PS appears to be counter intuitive. However, we note that the PS of a tracer field depends on two factors: the halo distribution with respect to the underlying matter (halo bias) and how the tracer populates the halos of varying properties (halo occupation). Even if the halo distribution becomes non-linear (Appendix \ref{ap:modPSBS}), it is possible for the tracers to occupy the halos in such a way that the tracer PS is linear to smaller scales compared to the matter or halo PS.  Many studies (see \cite{sarkar2016modelling, penin2018scale, wang2021breakdown, osinga2025lost}) show that the tendency of HI to occupy smaller, isolated halos \citep{papastergis2013clustering} leads to the 21-cm PS being well modeled by a constant linear bias parameter even up to $k \sim 2 \, h \, {\rm Mpc}^{-1}$ at $z=1$. Also, \cite{sarkar2019modelling} show that the tree-level BS is sufficient to model the simulated HI BS upto $k \sim 0.7 \, {\rm Mpc}^{-1}$ at $z=1$. Effective Field theory (EFTofLSS, \cite{mcdonald2009clustering, baumann2012cosmological, senatore2015ir, angulo2015statistics, ivanov2024effective, akitsu2025cosmology}) provides another alternative for modeling the 21-cm PS and BS, but we have not considered this here.} 

At the linear order, the PS is predicted to be  
\begin{equation}
    [P_\text{T}]_{\rm mod} (k)  = \left(\frac{4 \bar{T}}{3 \Omega_b}\right)^2 [\Omega_\text{HI} b_1]^2 P (k)
    \label{eq:ps}
\end{equation}
where $P(k)$ is the $\Lambda$CDM {linear} matter PS.  
Here the product $[\Omega_\text{HI} b_1]^2$ quantifies the amplitude of  
$ P_\text{T} (k)$ the 21-cm PS relative to $P(k)$ the {linear} matter PS.

Using second-order perturbation theory \citep{fry1984galaxy}, \cite{matarrese1997large} and \cite{scoccimarro2000bispectrum} have obtained the BS of a biased tracer of the underlying matter distribution. Here we use this to model the BS of the redshifted 21-cm brightness temperature fluctuations \citep{saiyad2006probing} as

\begin{equation}
    \begin{split}
        [B_\text{T}]_{\rm mod} ({k_1},{k_2},{k_3}) &= 2 \left(\frac{4 \bar{T}}{3 \Omega_b}\right)^3 [\Omega_\text{HI} b_1]^3 \left( F(\mathbf{k_1},\mathbf{k_2}) + \frac{\gamma}{2} \right)  \\ &  \times P (k_1) P (k_2) + \text{cyc..}
    \end{split}
    \label{eq:bs}
\end{equation}
where $\gamma = b_2/b_1$, and
\begin{equation}
    F(\mathbf{k_1},\mathbf{k_2}) = \frac{5}{7} + \frac{\mathbf{k_1} \cdot \mathbf{k_2}}{2} \left( \frac{1}{k_1^2} + \frac{1}{k_2^2} \right) + \frac{2}{7} \left( \frac{\mathbf{k_1} \cdot \mathbf{k_2}}{k_1 k_2} \right)^2
\end{equation}
are dimensionless kernels that depend on the shape of the triangle.
\cite{mazumdar2020quantifying} present simple analytic expressions for the kernels in terms of $(\mu,t)$ that parameterize the shape of the triangle.  Here the product $[\Omega_\text{HI} b_1]^3$ quantifies the amplitude of  
$ B_\text{T} ({k_1},{k_2},{k_3})$  relative to the matter BS 
\begin{equation}
        B ({k_1},{k_2},{k_3}) = 2  \left[  F(\mathbf{k_1},\mathbf{k_2}) P (k_1) P (k_2) + \text{cyc..} \right] \,.
    \label{eq:mbs}
\end{equation}
The quadratic bias parameter $\gamma$ influences the shape $(\mu,t)$  dependence of  $B_\text{T}(k_1,\mu,t)$ relative to $B(k_1,\mu,t)$. At large scales, we expect equations (\ref{eq:ps}) and (\ref{eq:bs}) to adequately model the 21-cm PS and BS, respectively. The model has two free parameters, namely $[\Omega_\text{HI} b_1]$ and $\gamma$, which depend on the HI abundance and how the HI is distributed relative to the underlying matter distribution, both of which are encoded in the HIHM relation. In principle, it is possible to jointly model the measured PS and BS, and use this to estimate  $[\Omega_\text{HI} b_1]$  and $\gamma$, and in turn use these to constrain the HIHM relation. Here, we demonstrate this using simulations.

\subsection{Estimating the bias parameters}
\label{subsec:bias mcmc}

In this section, we consider the possibility of estimating the bias parameters $[\Omega_\text{HI} b_1]$ and $\gamma$, using a measurement of the 21 cm PS and BS. 
We assume that the measured 21 cm PS and BS correspond to the  HIHM relation with the fiducial parameter values, and the errors in the measured values are entirely due to the cosmic variance arising from the finite simulation volume,  which are shown in Figures \ref{fig:ps}, \ref{fig:bs} and \ref{fig:bs_all}. The errors in a 21-cm intensity mapping experiment will typically have a noise contribution, in addition to the cosmic variance. However, the noise becomes subdominant, and can be neglected, given sufficiently deep observations. Our analysis here considers the best-case scenario where the noise is neglected, and the errors are entirely due to the cosmic variance. Here, we particularly focus on quantifying the precision to which it will be possible to estimate the parameters $[\Omega_\text{HI} b_1]$ and $\gamma$, given the measured 21-cm PS and BS with the errors as discussed above. 

We use equations (\ref{eq:ps}) and (\ref{eq:bs}) to model the 21-cm PS and BS respectively. 
For this, we assume that  $\bar{T}$ and $P(k)$ are known, as inferred using the $\Lambda$CDM parameters from \cite{refId0}, and we treat   $[\Omega_\text{HI} b_1]$ and $\gamma$  as free parameters.  Here, we determine the best fit values of $[\Omega_\text{HI} b_1]$ and $\gamma$ for which the model predictions best fits the measured 21-cm PS and BS. 
This is achieved by $\chi^2$ minimization, where 
\begin{equation}
    \label{eq:chisq}
    \chi^2 = (\mathbf{A} - [\mathbf{A}]_{\rm mod})^{T} \, \mathbf{C}^{-1} \, (\mathbf{A} - [\mathbf{A}]_{\rm mod})
\end{equation}
Here $\mathbf{A}$ is a column vector that includes all the data points of the measured PS and BS, and $[\mathbf{A}]_{\rm mod}$ includes their respective model predictions. We expect equations (\ref{eq:ps}) and (\ref{eq:bs}) to adequately model the observed (simulated here) data only at large scales,  and it is necessary to restrict the $k$ range to $k \le k_{ul}$ in eq.~(\ref{eq:chisq}) in order to avoid the small length scales where non-linear effects become important,  and we do not expect the perturbative analytic results to work. By trial and error, we find that the analytical models are in good agreement with the simulations for $k_{ul}= 0.32 \mbox{ Mpc}^{-1}$, and we have used this value for our analysis. At larger $k$ ($ > k_{ul}$),  the shape of the observed PS and BS differs from that of the analytic predictions, and the two cannot be reconciled by varying $[\Omega_\text{HI} b_1]$ and $\gamma$. The value of $k_{ul}$ is shown by the green vertical line in Fig.~\ref{fig:ps} and the top panel of Fig.~\ref{fig:bs}. { Note that for the contribution of the BS in eq.~(\ref{eq:chisq}),  we have taken all the points which are on the left of the green vertical lines corresponding to $k_{ul}$ in various $(\mu,t)$ planes of Fig.~\ref{fig:bs_all}. For this value of $k_{ul}$, we have $8$ values of the PS and $57$ values of the BS, whereby $\mathbf{A}$ has dimension $65$.
}

Here $\mathbf{C}$ is the $65 \times 65$ dimensional covariance matrix for the combined measurement of PS and BS whose diagonal elements correspond to the square of the errors on the data points in Figures \ref{fig:ps}, \ref{fig:bs}, and \ref{fig:bs_all}. The covariance matrix is predicted to be diagonal for a Gaussian random field. However, the 21-cm signal being considered here is non-Gaussian, which will give rise to off-diagonal terms in $\mathbf{C}$. While these off-diagonal terms are expected to be very important for the highly non-Gaussian EoR 21-cm signal \citep{mondal2015effect, mondal2016statistics, mondal2017statistics,shaw2019impact, shaw2020impact}, we do not expect these to be as significant for the mildly nonlinear post-EoR signal being considered here. The off-diagonal terms arise from the higher order statistics, namely the trispectrum for the PS \citep{mondal2015effect}, and the fifth and sixth order polyspectra for the BS (e.g. \citealt{sugiyama2020perturbation, mazumdar2020quantifying}).  While it is possible to analytically estimate these higher-order statistics for a biased tracer of the underlying matter distribution using higher-order perturbation theory, this is beyond the scope of the present work. 
{In this analysis, we have used the covariance matrix obtained from $100$ statistically independent realizations of the HI simulation, which includes the contribution from the correlations between the PS and BS at various k-values. 
Appendix \ref{ap:correlation} provides a visualisation of the dimensionless correlation matrix for the covariance matrix $\mathbf{C}$.
To assess the impact of finite-sample noise in the covariance, we performed a resampling analysis, wherein we repeatedly reconstructed the covariance from random subsets of the realizations and estimated the parameters $[\Omega_{\rm HI} b_1]$ and $\gamma$. We find that the scatter in the parameters is significantly smaller than their corresponding statistical uncertainties ($\sim 30$ per cent), demonstrating that the covariance estimated from $100$ realizations is 
sufficiently stable for the present likelihood analysis. Also, we have corrected the inverse covariance for the finite number of realizations using the Hartlap factor \citep{hartlap2007your} throughout our analysis.}


\begin{table}
	\centering
	\caption{Prior ranges, best-fit values and $1 \sigma$ errors for the estimated for $[\Omega_{\rm HI} b_1]$ and $\gamma$ obtained using the 21-cm PS and BS corresponding to fiducial parameters of HIHM relation.}
	\label{tab:bestfits}
	\begin{tabular}{lcc} 
        \hline
		Parameter & Prior & Best-fit\\
		\hline
		  $[\Omega_\text{HI} b_1]$  & $[0,1]$ & $(0.90 \pm {0.02}) \times 10^{-3}$\\\\
     	$\gamma$ & $[-5,5]$ & ${-0.36 \pm 0.06}$ \\
        \hline

	\end{tabular}
\end{table}

\begin{figure}
\centering
	\includegraphics[width=0.8\columnwidth]{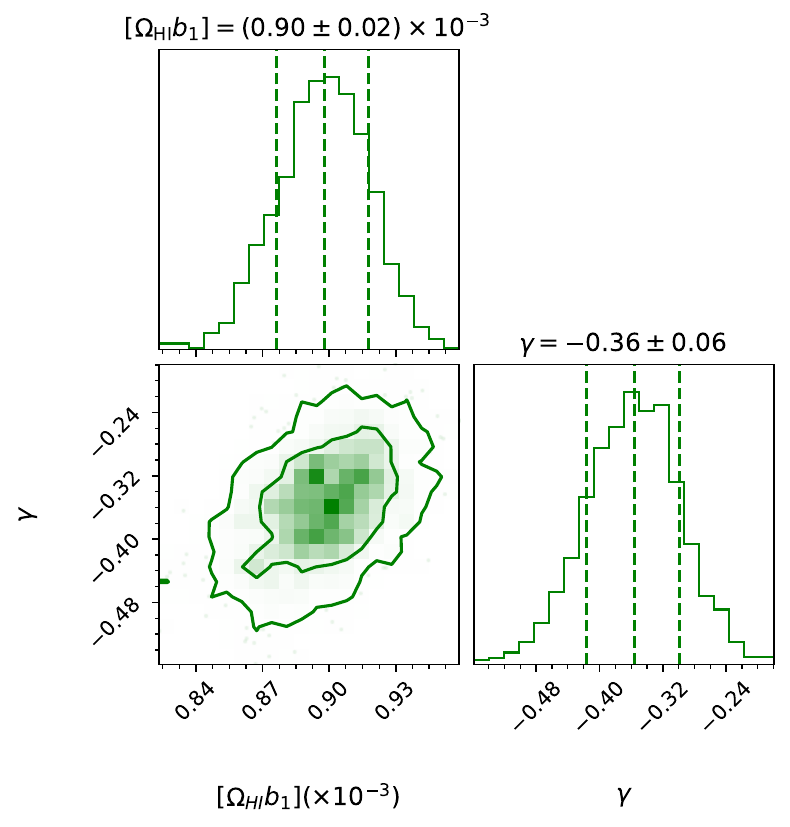}
    \caption{Estimates of $[\Omega_{\rm HI} b_1]$ and $\gamma$ obtained using the 21-cm PS and BS corresponding to fiducial parameters of HIHM relation. The contours show the $68$ percent and $95$ per cent confidence intervals. The histograms show the marginalized 1D posterior distributions for parameters. The three dashed lines in the 1D plots indicate the $16$th, $50$th and $84$th percentiles respectively.}
    \label{fig:bias_corner}
\end{figure}

We estimate the two unknown bias parameters $[\Omega_\text{HI} b_1]$ and $\gamma$ using an MCMC analysis with the \texttt{emcee} python package \citep{foreman2013emcee},  assuming flat priors for these parameters with ranges as indicated in Table \ref{tab:bestfits}. Here, we sample a Gaussian likelihood function using 16 random walkers for 1000 steps and discard the initial $10$ per cent samples as burn-in. The posterior distribution, shown in Fig.~\ref{fig:bias_corner}, exhibits tight constraints on both parameters. The 2D parameter plot shows the contours indicating $68$ per cent and $95$ per cent confidence intervals. We use the medians of the respective marginalized 1D distribution as the best-fit values of the parameters $[\Omega_{\rm HI}b_1]$ and $\gamma$.  The best-fit values and their $68$ per cent confidence intervals are  {$[\Omega_\text{HI} b_1]=(0.90 \pm {0.02} )\times 10^{-3}$  and $\gamma= -0.36 \pm {0.06}$}, which are also quoted in Table \ref{tab:bestfits}. The solid lines in Figs.~\ref{fig:ps}, \ref{fig:bs} and all the panels of Fig.~\ref{fig:bs_all} show the respective analytical predictions for the PS and BS  (eqs.~\ref{eq:ps} and \ref{eq:bs}) using the best-fit bias values of $[\Omega_{\rm HI}b_1]$ and $\gamma$. We find that the analytical predictions fit the simulated data for $k \leq k_{ul},$  whereas the simulated data exceed the analytical predictions at larger $k$. Finally, it is worth noting that the estimated value of $\gamma$ is negative for the fiducial values of the HIHM parameters. 

{We note here that the values of the parameters $[\Omega_{\rm HI} b_1]$ and $\gamma$ quoted above are obtained by fitting the predicted 21-cm PS (eq.~\ref{eq:ps}) and BS (eq.~\ref{eq:bs}) to the simulated 21-cm PS and BS respectively. Therefore, such estimates are, by construction, dependent on the details of the underlying HI simulation. However, in our analysis, these parameters act only as an intermediate link connecting the small-scale HI model to the large-scale 21-cm statistics, and are sufficiently robust for the purposes of this study.}

{
The halo model \citep{cooray2002halo} provides an alternative methodology.   
Instead of estimating $\Omega_{\rm HI}$ and $b_1$ by fitting the 21-cm PS and BS, it is also possible  to analytically predict  these parameters by employing the halo model formalism \citep{chen2021extracting}. However,  it is not straightforward to extend this 
to calculate $\gamma$, as we are not aware of a halo model framework for the BS.  It is therefore necessary to estimate $\gamma$ by fitting the predicted 21-cm BS to the simulated BS. We have considered this alternative approach in Section~\ref{subsec:halomod}.
}

\section{Estimating the HIHM parameters}
\label{sec:param est}

In this section, we consider the possibility of constraining the HIHM relation using measurements of 21 cm PS and BS at large length scales (small $k$). In particular, we consider the HIHM relation (eq. \ref{eq:HImodel}) given by \citet{padmanabhan2017halo} that has three parameters, namely $\alpha$, $\beta$, and $v_{c0}$, which respectively correspond to the normalization, the excess logarithmic slope and the lower cutoff on the circular velocity of haloes that may host HI. 
However, it is not possible to simultaneously constrain these three parameters using just the values of $[\Omega_{\rm HI}b_1]$ and $\gamma$ that completely quantify the large-scale 21 cm PS and BS.  This issue can be resolved by considering another independent observation. Here we consider $\Omega_{\rm HI}$, for which estimates already exist across a range of redshifts (e.g. \citealt{prochaska2005sdss, zwaan2005hipass, martin2010arecibo, freudling2010deep, zafar2013eso, rhee2018neutral}). In a recent work, \citet{chowdhury2020h} stacked the 21 cm emission at the positions of individual galaxies and measured $\Omega_{\rm HI} = (4.5 \pm 1.1) \times 10^{-4}$ at $z = 1.06$, which is close to the redshift of our interest. We note that this value is obtained by extrapolating the measurement from bright blue star-forming galaxies to the fainter limit. The contribution from red galaxies and other fainter sources, which are not included in this measurement, can only increase the value of $\Omega_{\rm HI}$. 
Our entire analysis here is based on \citet{padmanabhan2017halo} where the fiducial values of the HIHM parameters $(\alpha, \beta, v_{c0})$ have been tuned to match a wide variety of observations, and this predicts { $\Omega_{\rm HI} =1.12 \times 10^{-3}$ at $z = 1$ \citep{chen2021extracting}}, which is almost twice the measured value discussed above. 
We have used this value in our analysis, considering two cases which correspond to measurements with a relative accuracy $\Delta \Omega_{\rm HI}/\Omega_{\rm HI} $ of  $5 $ per cent and $1 $ per cent, respectively. 

\begin{figure*}
\centering
	\includegraphics[width=\textwidth]{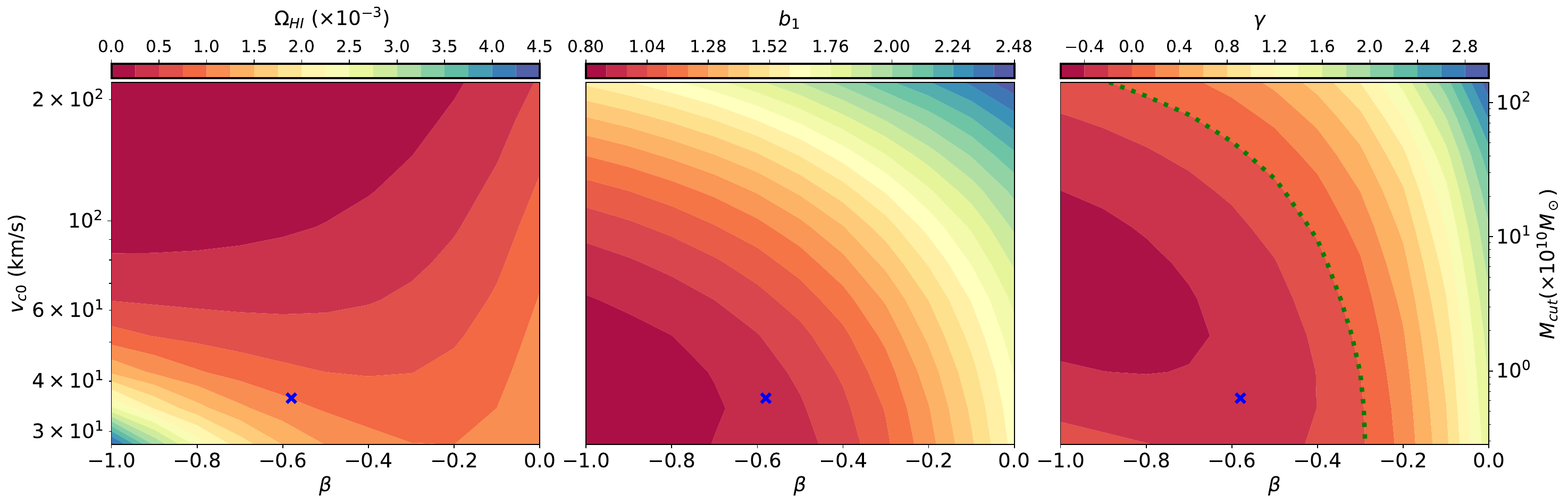} 
        
    \caption{The variation of $\Omega_\text{HI}$ ({Left}), $b_1$ ({Middle}), and $\gamma = b_2/b_1$ ({Right}) with HIHM parameters $\beta$ and $v_{c0}$. The corresponding variation of the mass cutoff at $z=1$ is also shown on the right vertical axis. The green dotted line in the right panel marks $\gamma = 0$. The blue cross in all panels shows the location of the fiducial parameters. The parameter $\alpha$ is chosen to be at the fiducial value.}
    \label{fig:contours}
\end{figure*}

We now analyze how the parameters $\Omega_{\rm HI}$, $b_1$ and $\gamma$, which can be estimated observationally, are related to $\alpha$, $\beta$ and $v_{c0}$, which are the parameters of the HIHM model. For this, we have simulated the HI distribution for a set of different HIHM parameter values in the range $0.01 \leq \alpha \leq 0.5$, $-1 \leq \beta \leq 0$, and $25 \leq v_{c,0} \leq 220 \, {\rm km \, s^{-1}}$, which translates to a lower mass cut-off range of $(0.2 - 139) \times 10^{10} M_\odot$ at $z=1$. For each case, we estimate the 21 cm PS and BS, and use these to estimate $[\Omega_{\rm HI} b_1]$ and $\gamma$. Throughout, we restrict the $k$ range to $k \leq k_{ul} = 0.32 \mbox{ Mpc}^{-1}$, and follow the methodology described in Section~\ref{sec:Bias est}. 
We  estimate $\Omega_{\rm HI}$ from the simulated HI distribution by summing over the HI contribution from every halo (labeled using $i$) 
\begin{equation}
    \label{eq:omega_HI}
    \Omega_{\rm HI} = \frac{1}{V \rho_{c0}} \sum_{i} M_{\rm HI}(i)  \,,
\end{equation}
where $V$ is the comoving volume of our simulation and $\rho_{c0}$ is the present value of the critical density. {The value of $ \Omega_{\rm HI}$ estimated this way will, in general,  differ from the analytical prediction based on the halo model \citep{chen2021extracting} due to the incompleteness of the simulated halos when the halo mass approaches the mass resolution of the simulation. In order to quantify the completeness of our simulations, we have compared the simulated value of $\Omega_{\rm HI}$ with the analytical prediction. We find that the for  the fiducial HIHM parameters, the simulated value is $\Omega_{\rm HI}=9.8 \times 10^{-4}$, which is $12.5 \%$ lower that the analytical prediction, and the under-estimate increases to $17 \%$ for the lowest value of the cut-off mass ($M_{\rm cut} \approx 2 \times 10^9 \, M_{\odot}$) used here.}

The left, middle and right panels of Fig.~\ref{fig:contours} show how $\Omega_{\rm HI}$, $b_1$ and $\gamma$ vary as functions of $\beta$ and $v_{c0}$. Note that the parameter $\alpha$ only scales the total HI content, which affects only $\Omega_{\rm HI}$, leaving $b_1$  and $\gamma$ unchanged, i.e., $b_1$  and $\gamma$ do not depend on $\alpha$. Furthermore, $\Omega_{\rm HI}$ scales linearly with $\alpha$ (eq.~\ref{eq:HImodel}), and it therefore suffices to show $\Omega_{\rm HI}$ only for the fiducial value of $\alpha$, as in the left panel of Fig.~\ref{fig:contours}.  The fiducial values of $\beta$ and $v_{c0}$ are marked by a blue cross in all the panels. Here $\beta$, the excess logarithmic slope,   determines how $M_{\rm HI}$ scales with $M_h$ (i.e. $M_{\rm HI} \propto M_h^{1+ \beta}$), and we have $M_{\rm HI} \propto M_h^{1}$, $M_{\rm HI} \propto \sqrt{M_h}$ and $M_{\rm HI}={\rm constant}$, respectively, for $\beta=0,-0.5$ and $-1$, which spans the range of values considered here. Furthermore, $v_{c0}$ the lower cutoff on circular velocity determines $M_{\rm cut}$ the lower halo mass cutoff, where the haloes with $M_h \lesssim M_{\rm cut}$ do not contain any HI. The values of $M_{\rm cut}$ corresponding to $v_{c0}$ are shown on the right vertical axis of each panel. 

The left panel of Fig.~\ref{fig:contours} shows $\Omega_\text{HI}$ as a function of $\beta$ and $v_{c0}$. We see that $\Omega_\text{HI}$ decreases monotonically with increasing $v_{c0}$. This is because the HI contribution of low mass haloes is removed as $M_{\rm cut}$ increases.  Considering the variation of $\Omega_\text{HI}$ with $\beta$, for small values of $M_{\rm cut}$, $\Omega_\text{HI}$ decreases with increasing $\beta$, while it is reversed for large $M_{\rm cut}$, where $\Omega_\text{HI}$ increases with increasing $\beta$. This transition occurs at around $M_{\rm cut} \approx 10^{11} M_\odot$, which is the pivot value in the HIHM relation (eq. \ref{eq:HImodel}).  The highest value of $\Omega_\text{HI} \approx 4.3 \times 10^{-3}$ lies in the bottom left corner ($v_{c0} \approx 30 \, {\rm km \, s^{-1}}$, $\beta \approx -1$), while $\Omega_\text{HI}$ is close to zero ($\sim 10^{-5}$) for a large part of our parameter space enclosed by the last contour in the upper left corner.

The middle panel shows the variation of the linear bias ($b_1$) with $\beta$ and $v_{c0}$. The lowest value ($b_1 \approx 0.83$) is found in the bottom left corner,  and $b_1$ increases monotonically with increasing $\beta$ and $M_{\rm cut}$. The highest value of $b_1 \, ( \approx 2.48)$ is found in the top right corner, where the signal is dominated by HI in the high mass haloes, which are known to be more tightly clustered as compared to smaller mass haloes \citep{sheth1999large}.

The right panel shows the variation of $\gamma = b_2/b_1$ with $\beta$ and $v_{c0}$.   The quadratic bias is rather interesting, as it can take both positive and negative values.
The green dotted line in the right panel marks $\gamma = 0$. We see that $\gamma$ assumes a negative value for more than half of our parameter space, including our fiducial value.
We find the minimum value of $\gamma \, (\approx -0.4)$ in the middle $(M_{\rm cut} \sim 10^{10} - 10^{11} \, M_{\odot})$   left $(\beta \lesssim -0.6)$ of the panel. Throughout, the value of $\gamma$ increases with increasing $\beta$, $\gamma$ is negative for small $\beta$ and positive for large $\beta$ where it becomes independent of $M_{\rm cut}$ as $\beta \rightarrow 0$. The general picture that emerges is that low mass haloes contribute to a negative $\gamma$, whereas high mass haloes contribute to making $\gamma$ positive. The final value of $\gamma$ depends on the relative HI contribution from low mass and high mass haloes, which changes as $\beta$ and $M_{\rm cut}$ are varied.

\begin{figure*}
\centering
	\includegraphics[width=0.45\textwidth]{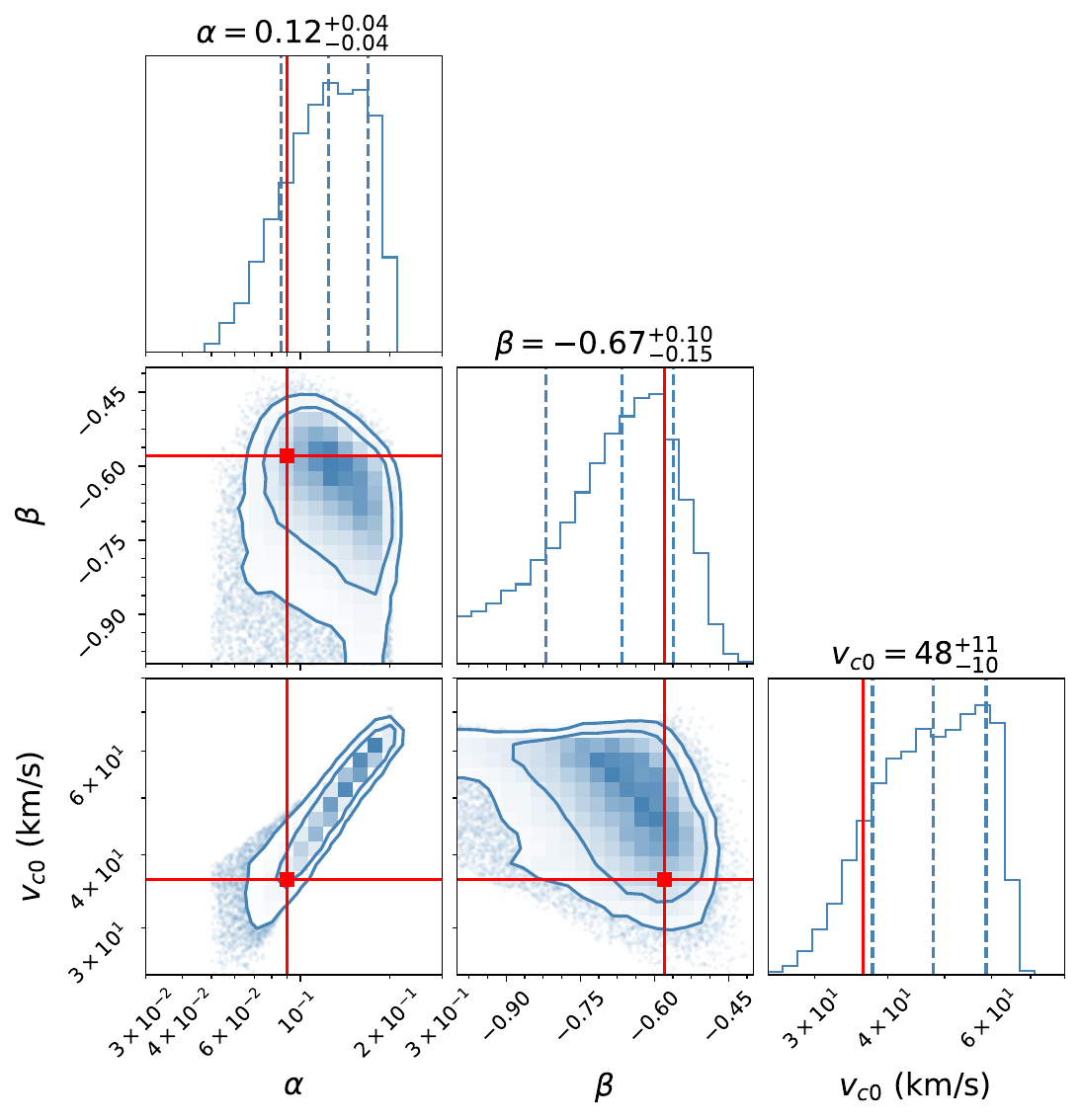} 
        \hspace{0.8cm}
        \includegraphics[width=0.45\textwidth]{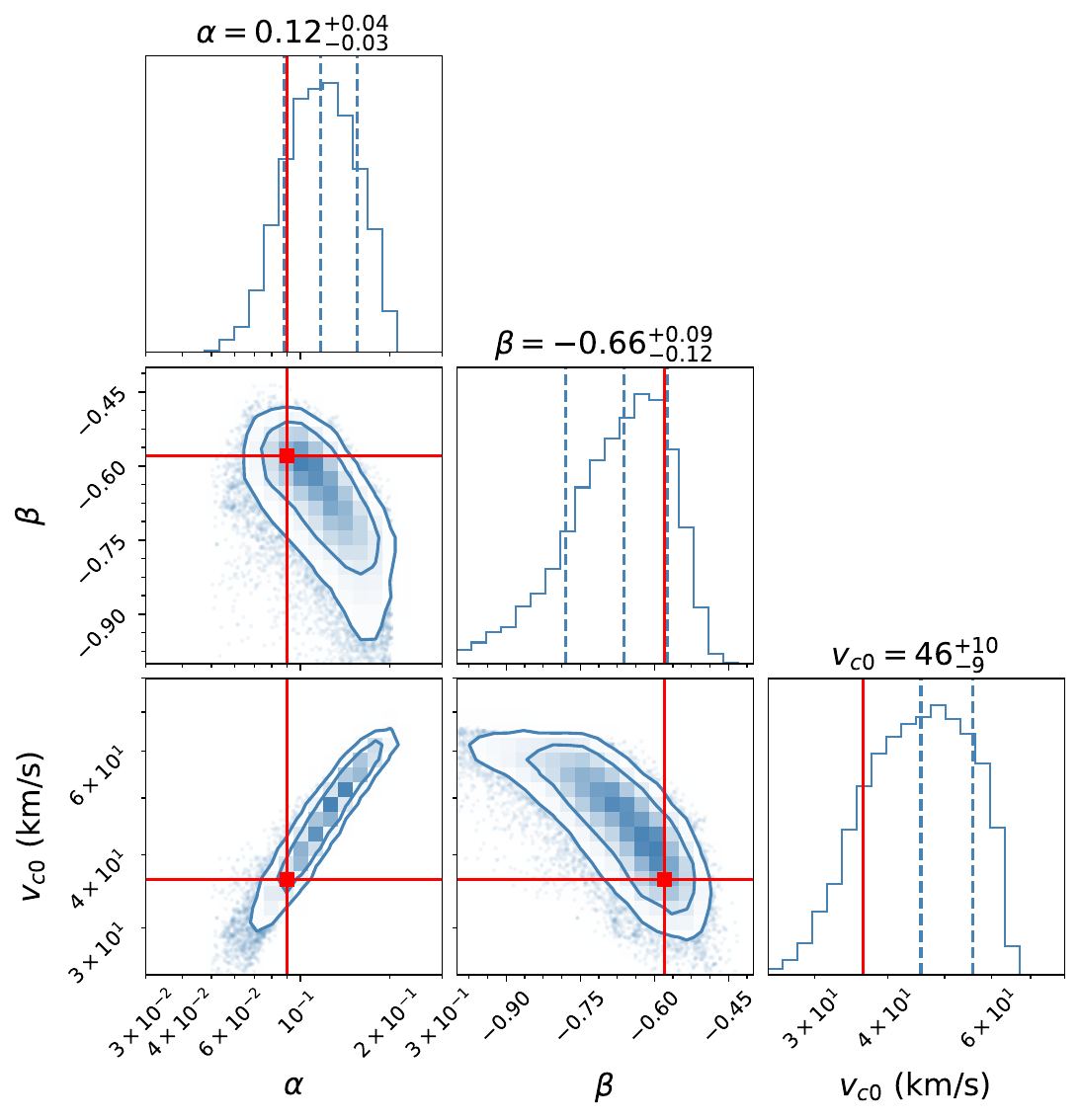}
    \caption{Constraints of the HIHM parameters, $\alpha$, $\beta$ and $v_{c0}$, using the redshifted 21 cm PS and BS along with an independent measurement of $\Omega_{\rm HI}$ with an assumed relative accuracy of $5 $ per cent (left panel) and $1 $ per cent (right panel). The contours show $68 $ per cent and $95 $ per cent confidence intervals in the joint distributions. The histograms show the marginalized 1D posterior distributions for parameters. The three dashed lines in the 1D plots indicate the $16$th, $50$th and $84$th percentiles respectively. The red points and red lines represent the fiducial values of the HIHM parameters used to simulate the "measured" signal.}
    \label{fig:param_corner}
\end{figure*}

Fig.~\ref{fig:contours} clearly shows that $\Omega_\text{HI}$, $b_1$ and $\gamma$, which quantify the statistics of the large-scale HI distribution, are all sensitive to $\beta$ and $v_{c0}$. Furthermore, $\Omega_\text{HI}$ also scales linearly with $\alpha$, which is not shown in the figure. It is therefore quite clear that a measurement of $\Omega_\text{HI}$, combined with measurements of the 21-cm PS and BS on large scale, can be used to constrain the values of $(\alpha,\beta,v_{c0})$ that parameterize the HIHM model and quantify how the HI is distributed among the haloes. 

As earlier, here also we assume a measured 21 cm PS and BS that correspond to the fiducial values of the HIHM parameters.  Furthermore,  the errors in these measurements are only due to cosmic variance, as shown in Figs.~\ref{fig:ps}, \ref{fig:bs} and \ref{fig:bs_all}.  As discussed in Section~\ref{subsec:bias mcmc}, the error estimates here represent the best-case scenario, and we usually will have larger errors in a more realistic situation.  In addition, we also consider an independent measurement of $\Omega_\text{HI}$,  considering two cases which correspond to measurements with a relative accuracy $\Delta \Omega_{\rm HI}/\Omega_{\rm HI} $ of  $5 $ per cent and $1 $ per cent respectively.  Given these measurements of the 21-cm PS and BS, and $\Omega_\text{HI}$, we quantify the precision to which it will be possible to constrain $(\alpha,\beta,v_{c0})$ the parameters of the HIHM relation. 
In other words, we estimate the values of  $\alpha,\beta,v_{c0}$ for which the predicted $[\Omega_{\rm HI}]_{\rm mod}$, $[P_\text{T}]_{\rm mod}$ and $[B_\text{T}]_{\rm mod}$ best fits the measured $\Omega_\text{HI}$, $P_{\rm T}$ and $B_{\rm T}$. 

We now discuss how we have calculated $[\Omega_{\rm HI}]_{\rm mod}$, $[P_\text{T}]_{\rm mod}$ and $[B_\text{T}]_{\rm mod}$  for arbitrary values of $\alpha,\beta,v_{c0}$. For this, we have interpolated the values of  $\Omega_{\rm HI}$, $b_1$ and  $\gamma$ shown in  Fig.~\ref{fig:contours}  as functions of $\beta$ and $v_{c0}$. While this is adequate for $b_1$ and  $\gamma$ which do not depend on $\alpha$, it only provides the value of $\Omega_{\rm HI}$ at $\alpha_{\rm fid}$ the fiducial value of $\alpha$. We scale this as $(\alpha/\alpha_{\rm fid})$ to calculate $[\Omega_{\rm HI}]_{\rm mod}$, and use this  and the interpolated values of  $b_1$ and  $\gamma$ in equations (\ref{eq:ps}) and (\ref{eq:bs}) to calculate $[P_\text{T}]_{\rm mod}$ and $[B_\text{T}]_{\rm mod}$.
We have estimated the best-fit values of $\alpha,\beta,v_{c0}$ by minimizing
\begin{equation}
    \label{eq:chisq-1}
    \chi_{\rm 1}^2 = \chi^2 + \left( \frac{\Omega_{\rm HI} - [\Omega_{\rm HI}]_\text{mod}}{\Delta \Omega_{\rm HI}} \right)^2
\end{equation}
where $\chi^2$ is same as in eq.~(\ref{eq:chisq}), and we have introduced an additional term to incorporate $\Omega_{\rm HI}$. 

We use an MCMC analysis, similar to that described in Section~\ref{sec:Bias est}, to estimate the best-fit HIHM parameters. We use a Gaussian likelihood function with flat priors (indicated in Table \ref{tab:bestfits_HI}) to sample 5000 steps of 50 random walkers and discard the initial $10 $ per cent samples as burn-in. The posterior distributions are shown in Fig.~\ref{fig:param_corner}, where the $68 $ per cent and $95 $ per cent confidence contours are shown for the 2D joint distributions. The histograms show the 1D marginalized distributions for all three HIHM parameters. The three dashed lines in the 1D plots indicate the $16$th, $50$th and $84$th percentiles respectively. Also, the red points and red lines mark the location of the fiducial HIHM parameter values that were used to simulate the "measured" signal.

\begin{table*}
	\centering
	\caption{Prior ranges, best-fit values and $1 \sigma$ errors for the estimated HIHM parameters for $\Delta \Omega_{\rm HI}/\Omega_{\rm HI} $ of  $5 $ per cent and $1 $ per cent respectively. The fiducial values of the HIHM parameters are $\alpha = 0.09$, $\beta = -0.58$ and $v_{c0} = 36.3 \, {\rm km \, s^{-1}}$ .}
	\label{tab:bestfits_HI}
	\begin{tabular}{lcccc} 
		\hline
		Parameter & Prior & Best-fit ($5 \%$) & Best-fit ($1 \%$) & Halo model\\
            \hline
            $\alpha$ & $[0.01,0.5]$ & {$0.12 \, ^{+0.04}_{-0.04}$} & {$0.12 \, ^{+0.04}_{-0.03}$} & {$0.08 \, ^{+0.05}_{-0.03}$}\\\\
		$\beta$ & $[-1,0]$ & {$-0.67 \, ^{+0.10}_{-0.15}$} & {$-0.66 \, ^{+0.09}_{-0.12}$} & {$-0.74 \, ^{+0.13}_{-0.16}$}\\\\
		$v_{c0} {\rm (km \, s^{-1})}$ & $[25,220]$ & {$48 \, ^{+11}_{-10}$} & {$46 \, ^{+10}_{-9}$ } & {$38 \, ^{+13}_{-7}$ }\\
            \hline
	\end{tabular}
\end{table*}

The left panel of Fig.~\ref{fig:param_corner} shows the constraints obtained by assuming $\Delta \Omega_{\rm HI}/\Omega_{\rm HI} = 5 $ per cent. Considering the three 2D plots, we see that the joint distribution of $(\alpha, v_{c0})$ is well constrained, having $68 $ per cent and $95 $ per cent contours that are closed within the parameter range considered here. However, for the distributions involving $\beta$, the $95 $ per cent contours do not close within the parameter range considered here. The fiducial values of the HIHM parameters, indicated by the red points, are all within the region enclosed by the $68 $ per cent confidence contour. Considering the marginalized 1D distributions, for each parameter, we consider the median to be the best fit value, and we use the 16th and 84th percentile to estimate the $68 $ per cent $(1 \sigma)$ confidence interval. The best fit values and the $1 \sigma$ errors for the HIHM parameters are tabulated in Table~\ref{tab:bestfits_HI}. The 1D marginalized distributions for $\alpha$ and $\beta$ have well-defined peaks,  and for both of these parameters, the fiducial value is within $\pm 1 \sigma$ of the best fit value. However, the marginalized distribution for $v_{c0}$ has a broad peak, and  $v_{c0}$ is relatively poorly constrained.  
We find that the fiducial value of $v_{c0}$ is $1.04 \, \times \, \sigma$ away from the best fit value. Repeated attempts with an increased number of samples yield no change in the distribution for $v_{c0}$.

The right panel of Fig.~\ref{fig:param_corner}  shows the constraints obtained by assuming $\Delta \Omega_{\rm HI}/\Omega_{\rm HI} = 1 $ per cent.  Compared with $\Delta \Omega_{\rm HI}/\Omega_{\rm HI} = 5 $ per cent, we see that the $68 $ per cent and $95 $ per cent confidence contours shrink for all the 2D joint distributions, and all the contours are now closed within the parameter range considered here.  In all cases, the fiducial parameter values lie within the region enclosed by the  $68 $ per cent confidence contour.  The elongated shape of the contours indicates a strong correlation between the parameters $(\alpha,v_{c0})$,  whereas $(\alpha,\beta)$ and $(\beta, v_{c0})$ are anti-correlated. 
Considering the 1D marginalized distributions, we see that they are narrower compared to $\Delta \Omega_{\rm HI}/\Omega_{\rm HI} = 5 $ per cent. 
The $1 \sigma$ errors shrink by $10 - 25$ percent for all parameters and the fiducial values are within $\pm 1 \sigma$ of the best fit value.

Our results indicate that it is possible to accurately estimate the parameters of the HIHM relation for both cases considered here. Comparing $\Delta \Omega_{\rm HI}/\Omega_{\rm HI} = 5 $ per cent with $\Delta \Omega_{\rm HI}/\Omega_{\rm HI} = 1 $ per cent, the errors shrink, and the best fit values of the parameters are closer to the fiducial parameter values that were used to simulate the "measured" 21-cm signal.  

\subsection{Halo model approach}
\label{subsec:halomod}

{
In this section, we consider an alternative halo model approach to estimate the HIHM parameters. The halo model formalism is based on the principle that all matter lies inside discrete dark matter halos of varying sizes, and the statistical properties of the tracers of matter distribution can be predicted by understanding the abundance and clustering of these halos (see \citealt{cooray2002halo} for a comprehensive review). This approach has been extensively used to model the galaxy distribution \citep{skibba2009halo, sinha2018towards}, the HI distribution \citep{padmanabhan2016constraining, wolz2016intensity, padmanabhan2017halo, padmanabhan2017constraints}, and the IM signal of lines other than the 21-cm \citep{schaan2021multi}.
}

{
Here, we calculate the values of  $\Omega_{\rm HI}$ and $b_1$ analytically using the halo model formalism. We calculate $\Omega_{\rm HI}$ using eq.~(\ref{eq:omega_HI}), and similarly calculate $ b_1$ using  
\begin{equation}
    \label{eq:b1_hm}
    b_1 = \frac{\sum_i M_{\rm HI} (i) \, b_h (i)}{\sum_i M_{\rm HI} (i)} \, ,
\end{equation} 
where $b_h (M)$ is the linear halo bias for halos of mass $M$, which we have calculated using the analytical expressions given in \citet{sheth2001ellipsoidal}. We have used the average value of the parameters obtained from the $100$ realizations of the simulations. Instead of summing over the halos actually present in our simulations,  we can also calculate these parameters using a fully analytical approach by integrating over the halo mass function \citep{press1974formation, sheth1999large} with a lower mass limit to match our simulations.
We find that this yields a value of $[\Omega_{\rm HI} b_1]$ that is $12$ percent larger than the value $[\Omega_{\rm HI} b_1] = 0.94 \times 10^{-3}$ obtained from the summations. This may be because our simulations miss some of the halos near the lower mass limit, which is close to the mass resolution of our simulations. For the present analysis, we use the summation as this directly pertains to the simulations used to estimate the  21-cm PS and BS. For the fiducial HIHM parameters, the 21-cm PS  predicted (eq.~\ref{eq:ps})  using the value of $[\Omega_{\rm HI} b_1]$ obtained from the summation (eq.~\ref{eq:omega_HI} and \ref{eq:b1_hm}) lies well within the $2\sigma$ fluctuations of the simulated 21-cm PS for $k<0.32 \, {\rm Mpc}^{-1}$.  Furthermore, this value of $[\Omega_{\rm HI} b_1]$ is only $4$ percent larger than the one quoted in Table \ref{tab:bestfits}, which is obtained by fitting the simulated 21-cm PS and BS.
}

{
We are not aware of a halo model framework for $\gamma$ and the BS. It is therefore necessary to estimate $\gamma$ by fitting the predicted 21-cm BS (eq.~\ref{eq:bs}) to the simulated BS. We model the simulated 21-cm BS at large scales (i.e. $k < 0.32 \, {\rm Mpc}^{-1}$) using eq.~(\ref{eq:bs}), keeping $[\Omega_{\rm HI} b_1]$ fixed at the value obtained using equations (\ref{eq:omega_HI}) and (\ref{eq:b1_hm}). To estimate $\gamma$, we use an MCMC analysis, similar to the one described in Section~\ref{subsec:bias mcmc}. We sample 1000 steps of 16 random walkers from a Gaussian likelihood function and discard the first $10$ percent samples as burn-in. The best-fit value of $\gamma$ is chosen to be the median of the posterior distribution. For the fiducial HIHM parameters, the estimated value $\gamma = -0.44 \pm 0.06$ is $\sim 22$ percent larger than the best-fit value obtained in Section~\ref{subsec:bias mcmc} (Table \ref{tab:bestfits}). 
}
\begin{figure}
\centering
	\includegraphics[width=\columnwidth]{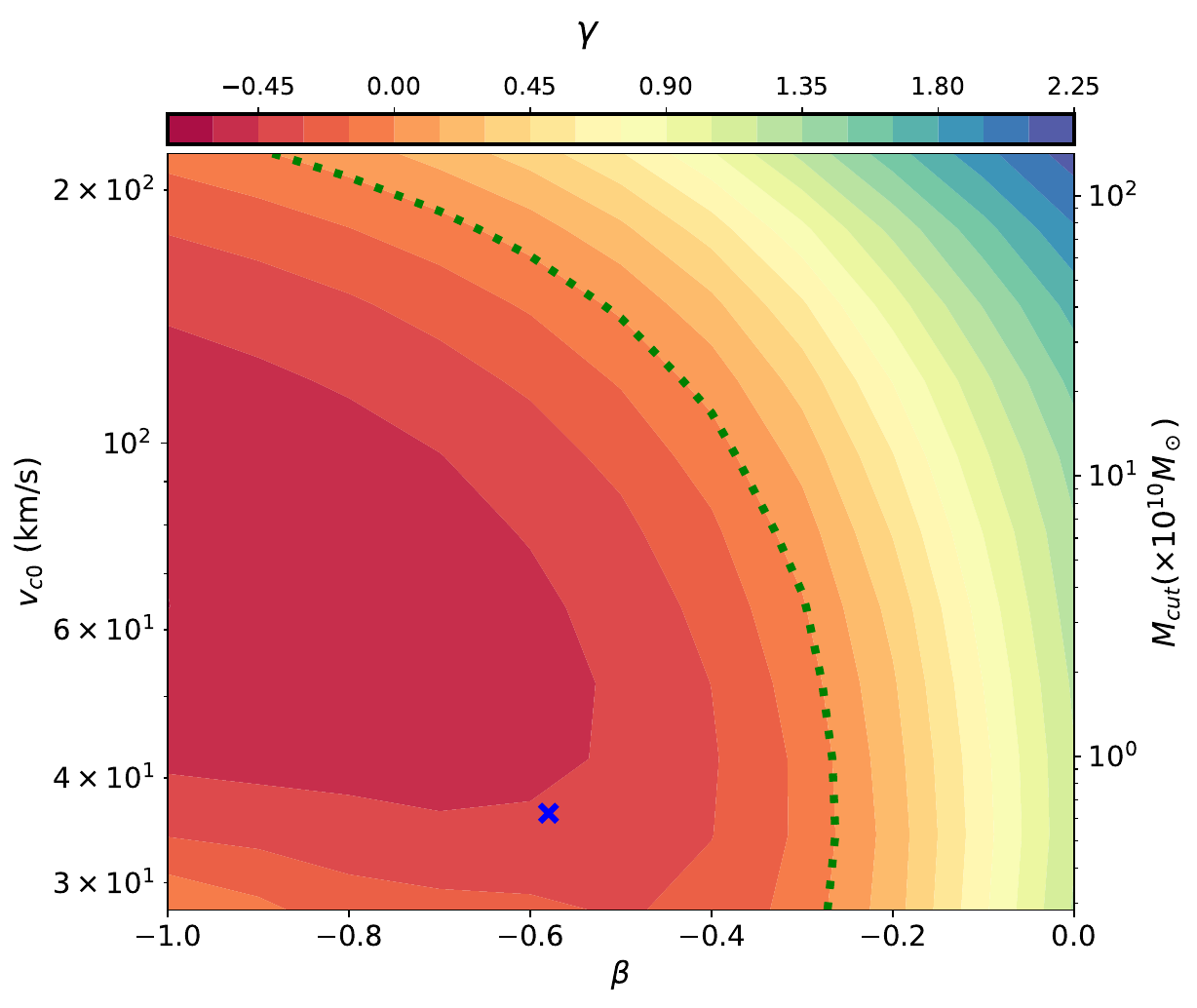}
    \caption{The variation in $\gamma$ as function of HIHM parameters $\beta$ and $v_{c0}$ estimated using the halo model approach. The $\gamma = 0$ contour is shown by the green dotted line and the blue cross marks the location of the fiducial values of $\beta$ and $v_{c0}$. Note that $\gamma$ is independent of $\alpha$.}
    \label{fig:gamma_grid}
\end{figure}

\begin{figure}
\centering
	\includegraphics[width=0.9\columnwidth]{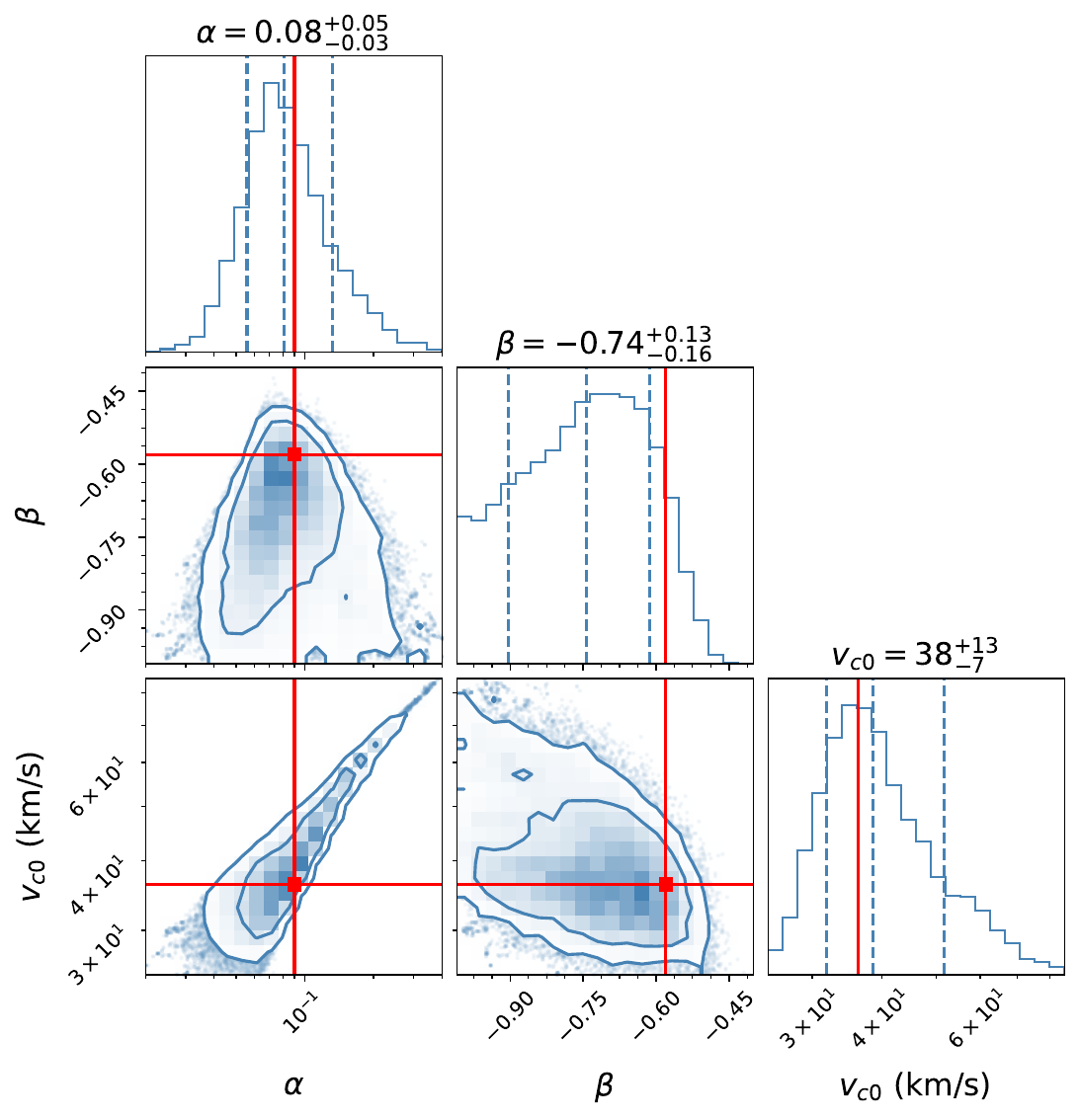}
    \caption{Constraints on the HIHM parameters $\alpha$, $\beta$ and $v_{c0}$ obtained using the halo model approach. The fiducial values of the parameters used to simulated the "measured" signal are shown by the red points and lines.}
    \label{fig:params_bs_hm}
\end{figure}

{
We now consider the possibility of estimating the HIHM parameters using the measurements described in Section~\ref{sec:param est} with the halo model approach. Here, we calculate $\Omega_{\rm HI}$ and $b_1$ as functions of the HIHM parameters $\alpha$, $\beta$, and $v_{c0}$ using equations (\ref{eq:omega_HI}) and (\ref{eq:b1_hm}) respectively. However, $\gamma$ is calculated by interpolating over the $\beta$ and $v_{c0}$ grid shown in Fig.~\ref{fig:gamma_grid}. Compared to the right panel of Fig.~\ref{fig:contours}, we see that the shapes of the contours remain unchanged throughout the parameter space, albeit we have an overall shift towards smaller values. We perform an MCMC analysis similar to the one described in the previous section using the measurements of the 21-cm BS and one independent measurement of $\Omega_{\rm HI}$, assuming $\Delta \Omega_{\rm HI}/\Omega_{\rm HI} = 1$ per cent. Fig.~\ref{fig:params_bs_hm} shows the posterior distribution for the HIHM parameters, where the $68$ per cent confidence contours are closed within the parameter range. However, for the distributions involving $\beta$, the $95$ percent contours do not close. The fiducial HIHM parameter values (indicated by red points) lie within the $68$ percent confidence interval of the 2D distribution. The best-fit values are taken to be the medians of the 1D marginalized distributions of the parameters, which are indicated in Table \ref{tab:bestfits_HI}. We find that the fiducial values of $\alpha$ and $v_{c0}$ lie within $68$ per cent confidence intervals around the best-fit values. The fiducial $\beta$, however, lies $1.2 \times \sigma$ away from the best-fit value. Comparing with the right panel of Fig.~\ref{fig:param_corner}, we find that the posterior distributions obtained using the halo model approach are broader and $1\sigma$ errors on the best-fit values are increased upto $\sim 30$ per cent.
}

{
Our results indicate that it is possible to estimate the HIHM parameters using the halo model approach by analytically calculating the $\Omega_{\rm HI}$ and $b_1$. However, since we do not have a reliable way to calculate the BS directly using the halo model, we have to estimate $\gamma$ by fitting the simulated BS. Furthermore, in a subsequent work, we would like to extend the present analysis to redshift space. While there exists a clear perturbative framework for modeling the redshift space PS and BS (eg. \cite{mazumdar2020quantifying}), we are not aware of a halo-based formalism that can be used to model the redshift space BS. 
}

\section{Summary and Discussion}
\label{sec:summary}

In the first part of this paper, we have simulated the $z=1$ Post-EoR 21-cm signal starting from a suite of $10$ dark matter only N-body simulations of comoving volume $[150.08 \mbox{ Mpc}]^3$ that we use to identify haloes. These haloes were populated with HI using the HIHM relation proposed by \citet{padmanabhan2017halo} that has three parameters  (eq.~\ref{eq:HImodel}), namely $\alpha$, $\beta$, and $v_{c0}$, which respectively correspond to the normalization, the excess logarithmic slope and the lower cutoff on the circular velocity of haloes that may host HI. The simulations and all the results presented here correspond to the fiducial values of the parameters, namely  $\alpha = 0.09$, $\beta = -0.58$, and $v_{c0} = 36.3 \, {\rm km \, s^{-1}}$. The simulated signal here does not include RSD, and we propose to address this in future work. 

We have quantified the statistics of the Post-EoR 21-cm signal using the 21-cm PS and BS. We present results for the 21-cm PS (Figure \ref{fig:ps}) covering the entire available $k$ range, and the 21-cm BS (Figure \ref{fig:bs} and \ref{fig:bs_all})  covering the entire $k_1$ range and all possible triangle shapes $(\mu,t)$.  Assuming that the 21-cm signal traces the underlying dark matter with a bias, we model the 21-cm PS and BS using linear perturbation theory (eq.~\ref{eq:ps}) and second order perturbation theory (eq.~\ref{eq:bs}) respectively. The model has two free parameters namely $[\Omega_{\rm HI} b_1]$ and $\gamma=b_2/b_1$. We find that the model predictions provide a good fit to the simulated 21-cm PS and BS, provided we restrict the comparison to large length scales $k \le k_{ul}= 0.32 \mbox{ Mpc}^{-1}$. 

We have considered the possibility of estimating the bias parameters $[\Omega_\text{HI} b_1]$ and $\gamma$, using a measurement of the 21 cm PS and BS. We assume that the measured PS and BS correspond to the  HIHM relation with the fiducial parameter values, and the errors in the measured values are entirely due to the cosmic variance. Our analysis, which ignores the noise contribution, represents the best-case scenario. Comparing the model predictions with the "measured" 21-cm PS and BS, we obtain the best-fit values {$[\Omega_\text{HI} b_1]=(0.90 \pm {0.02} )\times 10^{-3}$  and $\gamma= -0.36 \pm {0.06}$} (Table \ref{tab:bestfits}).


Finally, we consider the possibility of estimating the HIHM relation using measurements of the 21-cm PS and BS.  It is necessary to consider a third measurement to estimate the values of  $(\alpha,\beta,v_{c0})$, and here we consider an independent measurement of $\Omega_\text{HI}$. We consider two scenarios that respectively correspond to measurements with $5 $ per cent and $1 $ per cent relative accuracy.  We proceed by numerically modeling  $\Omega_{\rm HI}$, $b_1$ and $\gamma$ as functions of $\alpha,\beta,v_{c0}$, which allows us to model the 21-cm PS and BS as functions of $(\alpha,\beta,v_{c0})$. We use this to estimate the parameter values $(\alpha,\beta,v_{c0})$ for which the modeled 21-cm PS and BS best fit the "measured" 21-cm PS and BS. We find that the best-fit parameter values (Table~\ref{tab:bestfits_HI}) are consistent with the fiducial values that were used to simulate the 21-cm signal. Comparing the scenario with $\Delta \Omega_{\rm HI}/ \Omega_{\rm HI}=1 $ per cent with $\Delta \Omega_{\rm HI}/ \Omega_{\rm HI}=5 $ per cent, the $1 \sigma$ errors 
reduce by $10 - 25$ percent for all parameters.
The best fit values also move closer to the fiducial values used to simulate the signal. 

{In an alternative approach, we have considered the halo model formalism to predict the parameters $\Omega_{\rm HI}$ and $b_1$. We are not aware of any halo model based approach to predict $\gamma$, and we have to estimate $\gamma$ by fitting the simulated 21-cm BS.
We show that the HIHM parameters can be reliably constrained by using the measurements the 21-cm BS along with an independent measurement of $\Omega_{\rm HI}$ with $\Delta \Omega_{\rm HI}/ \Omega_{\rm HI}=1 $. Comparing the HIHM estimates obtained using the previous approach to the halo-model approach, we find that $1\sigma$ errors of the HIHM parameters are increased upto $\sim 30$ per cent in the latter scenario.
}

{The entire analysis presented here is limited by the fact that our methodology relies on parameters that have been estimated from simulations. These simulations have a limited mass resolution, and it misses the HI in halos below the mass limit. It is therefore possible that this may introduce a bias in the estimated HIHM parameters when we apply this methodology to the actual observational data. While the  halo model offers a possibility of overcoming this limitation by analytically evaluating the model parameters, at present its utility is limited as we are not aware of an analytical prescription for calculating $\gamma$. It is possible to quantify  the impact of the limited mass resolution by considering simulations with higher resolution, which can be used to test the convergence of our model parameters. We plan to address this in future work.   
}

The HIHM relation quantifies how the HI mass contained in a dark matter halo varies with the halo mass.  Here, we have shown that measurements of 21-cm PS and BS on large length scales  $(k \le k_{ul}= 0.32 \mbox{ Mpc}^{-1})$,  combined with an independent measurement of $\Omega_{\rm HI}$, can be used to estimate the parameters of the HIHM relation that quantifies how the HI is distributed at small scales. The HIHM relation tells us about the HI content of individual galaxies and groups of galaxies. We expect this to provide insights into various aspects of galaxy formation and evolution, particularly the processes like star formation that directly affect the state of the inter-stellar medium (ISM). The present study also demonstrates that large scale 21-cm PS and BS are well modeled using simple analytical prescriptions. We expect this to be important for studies that propose to use 21-cm intensity mapping for cosmological studies, such as parameter estimation and constraining primordial non-Gaussianity. 

The present work is intended to be a proof of concept that it is possible to estimate the parameters of the HIHM relation using measurements of the 21-cm PS and BS at large length scales. The entire analysis here has ignored the effect of peculiar velocities (RSD) on the 21-cm signal \citep{bharadwaj2004cosmic}.  A combined analysis of the 21-cm PS and BS, including RSD, has the potential to directly provide estimates of $\Omega_{\rm HI}$, $b_1$ and $\gamma$, 
without the need for an independent measurement of $\Omega_{\rm HI}$, which we plan to address in future work. Furthermore, we have assumed cosmic variance to be the only source of errors in the measured 21-cm PS and BS. This represents the best case scenario, as the errors will typically be larger due to the noise contribution which has been ignored here. In addition, some of the ${\bf k}$ modes may be inaccessible due to foreground contamination. We plan to address these issues to make more realistic predictions for existing and future telescopes like MeerKAT and SKA-Mid. We also plan to extend the analysis to cover other redshifts in the Post-EoR era. 

\section*{Acknowledgments}
We thank Debanjan Sarkar for providing the halo catalogues for the five realizations used in this work and also for useful discussions. MC acknowledges the support of the Prime Minister Research Fellowship (PMRF). We acknowledge the computing facilities in the Department of Physics, IIT Kharagpur.

\section*{Data Availability}
The simulated data and the package involved in this work will be shared on reasonable request to the corresponding author.

\bibliographystyle{mnras}
\bibliography{refer_old}

\appendix

\section{Bispectrum for all triangle configurations}
\label{ap:full bisp}
\begin{figure*}
\centering
	\includegraphics[width=\textwidth]{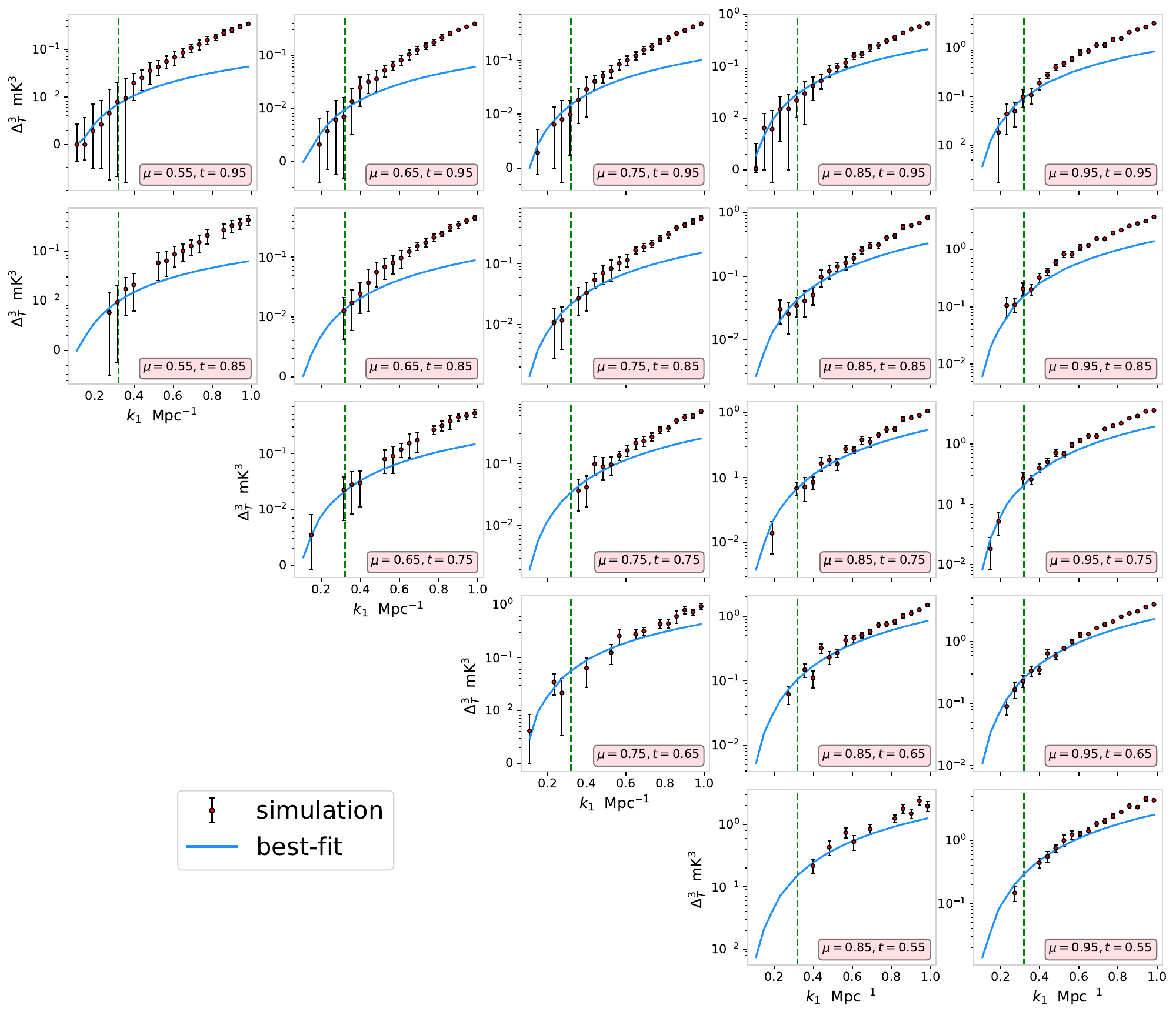}
    \caption{Mean-cubed 21 cm brightness temperature fluctuations ($\Delta_T^3$) for all possible triangle shapes. Each panel corresponds to a unique shape quantified by parameters $\mu$ and $t$ for which the bispectrum is shown as a function of size $k_1$. The red points are obtained from 100 independent realizations of the HI simulation described in Section \ref{sec:HI model}. The blue solid line shows the predicted model of the bispectrum (eq. \ref{eq:bs}) calculated using best-fit bias values of $[\Omega_{\rm HI} b_1]$ and $\gamma$ (Section~\ref{subsec:bias-model}). The green dashed line shows the extent of our fit ($k_{ul} = 0.32 \, {\rm Mpc}^{-1}$).}
    \label{fig:bs_all}
\end{figure*}

Here, we present the 21-cm bispectrum (BS) corresponding to the fiducial values of the HIHM parameters. Fig.~\ref{fig:bs_all} shows the mean-cubed brightness temperature fluctuations $(\Delta_{\rm T}^3)$ in a binned $\mu-t$ plane. Each possible triangle shape of a given size ($k_1$) can be uniquely mapped to a value of the shape parameters $(\mu,t)$ in the range $0.5 \leq \mu, t \leq 1$, with the constraint $2\mu t \leq 1$. As we move from left to right across the panels, $\mu$ changes from $0.55$ to $0.95$, while $t$ increases from bottom to top in the same range. The top left panel illustrates the BS of triangles that are closest to being equilateral (refer to the top panel of Fig.~\ref{fig:bs}). The top row, corresponding to $t = 0.95$, displays the behavior of L-isosceles triangles ($k_1 = k_2$) as they transition from equilateral to the squeezed limit in the top right corner. All panels along the lower left edge represent S-isosceles triangles ($k_2 = k_3$). The rightmost column highlights linear or flattened triangles, ranging from stretched to squeezed shapes. For more information on the allowed $\mu-t$ space, refer to fig.~2 of \cite{bharadwaj2020quantifying}. We show the same $k_1$ range $0.11 - 0.98 \mbox{ Mpc}^{-1}$ for all panels.

The red points with error bars are obtained from the simulated HI distribution described in Section~\ref{sec:HI model} using 100 independent realizations. We see that the overall BS increases with $k_1$ in a similar fashion for all shapes. It is consistent with zero for $k_1 \leq 0.2 \mbox{ Mpc}^{-1}$ for all panels and then increases to $\sim 0.4 \, {\rm mK}^3$ near the equilateral limit and $\sim 5 \, {\rm mK}^3$ near the stretched triangle limit. BS for linear triangles $\mu = 0.95$ does not change much with in $t$. However, as $\mu$ increases, the BS increases monotonically for both L- and S-isosceles triangles. The blue solid line represents the predicted BS (eq. \ref{eq:bs}) calculated using the best-fit values of $[\Omega_{\rm HI} b_1]$ and $\gamma$. The best-fit parameters are estimated using the combined 21-cm PS and BS on large scales (Section~\ref{subsec:bias mcmc}). The green dashed line $(k \leq 0.32 \mbox{ Mpc}^{-1})$ shows the extent of our fit, such that all the points on the left of this line are used for fitting. We find that our model is a good fit to the simulated HI BS at large length scales for all triangle shapes. However, as we go to smaller $k$ values, the predicted BS falls below the simulated BS by about an order of magnitude. 

\section{Modelling the PS and BS}
\label{ap:modPSBS}
In this section, we use the perturbative bias prescription \citep{fry1993biasing, desjacques2018large} along with Standard Perturbation Theory (SPT, \cite{goroff1986coupling, jain1993second, bernardeau2002large}) to model the PS and BS of a biased tracer of the underlying matter density field. We restrict our analysis to perturbative correction terms that are quadratic in the linear matter power spectrum, which corresponds to calculating the PS up to 1-loop order and the BS up to tree-level.

The tracer overdensity is modeled as a function of the local overdensity of the underlying matter field using a bias prescription up to the third order \citep{desjacques2018large},
\begin{equation}
    \delta_{\rm tr} (\mathbf{x}) = b_1 \delta (\mathbf{x}) + \frac{b_2}{2} \delta^2 (\mathbf{x}) + \frac{b_3}{6} \delta^3 (\mathbf{x})\, ,
    \label{eq:bias_3}
\end{equation}
where $b_1, \, b_2$ and $b_3$ are the linear, quadratic, and cubic bias parameters, respectively. 
The tracer overdensity in the Fourier space can be written as 
\begin{equation}
    \label{eq:deltaf}
    \begin{aligned}
    {}& \Delta_{\rm tr}  (\mathbf{k}) = \,  
    b_1 \Delta_1 (\mathbf{k}) + b_1 \int \frac{d^3 q}{(2 \pi)^3} \left[ F_2 (\mathbf{q}, \mathbf{k} - \mathbf{q}) + \frac{\gamma_2}{2} \right] \times \\ 
    & \Delta_1 (\mathbf{q}) \Delta_1 (\mathbf{k} - \mathbf{q})
     + b_1 \int \frac{d^3 q_1}{(2 \pi)^3} \frac{d^3 q_1}{(2 \pi)^3} \left[ \frac{}{} F_3(\mathbf{q_1}, \mathbf{q_2}, \mathbf{k} - \mathbf{q_1} - \mathbf{q_2}) \, + \right. \\
     & \left.  \gamma_2 F_2(\mathbf{q_2}, \mathbf{k} - \mathbf{q_1} - \mathbf{q_2}) + \frac{\gamma_3}{6} \right] \Delta_1 (\mathbf{q_1}) \Delta_1 (\mathbf{q_2}) \Delta_1 (\mathbf{k} - \mathbf{q_1} - \mathbf{q_2})
    \end{aligned}
\end{equation}
where $\gamma_2 = b_2/b_1$, $\gamma_3 = b_3/b_1$, and  $F_2$ and $F_3$ are the second and third-order SPT kernels, respectively \citep{goroff1986coupling}. The tracer overdensity at a given mode $\mathbf{k}$ receives contributions from the coupling of all the modes that add up to $\mathbf{k}$, i.e., $\mathbf{q_{\rm 1}} + \mathbf{q_{\rm 2}} + ... + \mathbf{q_{\rm N}} = \mathbf{k}$. Here, we only consider the contributions up to $N = 3$. 
The tracer PS up to the 1-loop order, calculated using eq. (\ref{eq:deltaf}), 
is predicted to be \citep{makino1992analytic, carlson2009critical, nishizawa2013perturbation}, 
\begin{equation}
    \label{eq:Ptr}
    \begin{aligned}
    P_{\rm 1-loop}  \, {}& (k) =
    b_1^2 \left\{ \,  \left[ 1 + \left( \frac{68}{21} \gamma_2 + \gamma_3 \right) \sigma^2 \right] P(k) \right. \\
     & \left. + \, 2 \int \frac{d^3 q}{(2 \pi)^3} \left[ F_2^{(s)} (\mathbf{q}, \mathbf{k} - \mathbf{q}) + \frac{\gamma_2}{2} \right]^2  P(q) P(|\mathbf{k} - \mathbf{q}|) \right. \\
     & \left. + \, 6 \, P(k) \int \frac{d^3 q}{(2 \pi)^3} F_3^{(s)} (\mathbf{k}, \mathbf{q}, - \mathbf{q}) P(q) \,  \right\} + 2 \pi^3 b_1^2 \gamma_2^2 \sigma^4
    \end{aligned}
\end{equation}
where the superscripts $(s)$ denote the symmetrized version of the SPT kernels \citep{goroff1986coupling}. Also, $P(k)$ is the $\Lambda$CDM linear matter PS, and $\sigma^2$ is the variance of the underlying matter field which can be calculated from  $P(k)$ using 
\begin{equation}
    \sigma^2 = \int \frac{d^3 q}{(2 \pi)^3} P(q) \, .
\end{equation}
Here, we limit the integration over $q$ by the box size ($\approx 10^{-2} \, {\rm Mpc}^{-1}$) and the spatial resolution ($\approx 10^{1} \, {\rm Mpc}^{-1}$) of our simulation, on the lower and upper ends respectively.
The tree-level BS for the tracer can be modelled using a similar perturbative expansion \citep{matarrese1997large, scoccimarro2000bispectrum},
\begin{equation}
    \label{eq:Btr}
    B_{\rm tree} (k_1, k_2, k_3) = 2 b_1^3\left[ F_2 (\mathbf{k_1}, \mathbf{k_2}) + \frac{\gamma_2}{2} \right] P(k_1)P(k_2) + {\rm cyc.}
\end{equation}

This model has three free parameters, namely $b_1, \, \gamma_2 \,  {\rm and} \, \gamma_3$, which quantify the linear, quadratic and cubic bias parameters respectively.   
We expect equations (\ref{eq:Ptr}) and (\ref{eq:Btr}) to respectively model the PS and BS of the simulated tracer distributions in the mildly nonlinear regime, where linear perturbation theory breaks down. However, the perturbative approach itself, including higher order terms, is restricted to a limited range of $k$ modes once non-linearity sets in and there is significant shell-crossing \citep{bharadwaj1996evolution}.  

We now compare this model (eq \ref{eq:Ptr} and \ref{eq:Btr}) to the simulated matter PS and BS at $z=1$. The bias parameters $(b_1, \, \gamma_2, \, \gamma_3)$ are set to $(1,0,0)$ for the unbiased matter distribution. We find that the linear PS $P(k)$ matches the simulated PS up to $k < 0.1 \, {\rm Mpc}^{-1}$, but fails at larger $k$.  The 1-loop PS (eq. \ref{eq:Ptr}) matches the simulated PS  within $\pm 2\sigma$ up to $k < 0.2 \, {\rm Mpc}^{-1}$. This is consistent with previous studies \citep{jeong2006perturbation, sefusatti2007bispectrum} that validate the 1-loop PS over a similar $k$ range at $z=1$. Considering the simulated BS, we find that this is   well modeled, within $2 \sigma$, by the tree-level BS (eq.~\ref{eq:Btr}) for different triangle shapes at $k_1 < 0.2 \, {\rm Mpc}^{-1}$. \footnote{The match is not very good at the squeezed limit for the smallest $k$-bin, most probably because of the limitation of the BS estimator arising from the finite bin size \citep{shaw2021fast}.}


\begin{figure*}
\centering
	\includegraphics[width=0.325\textwidth]{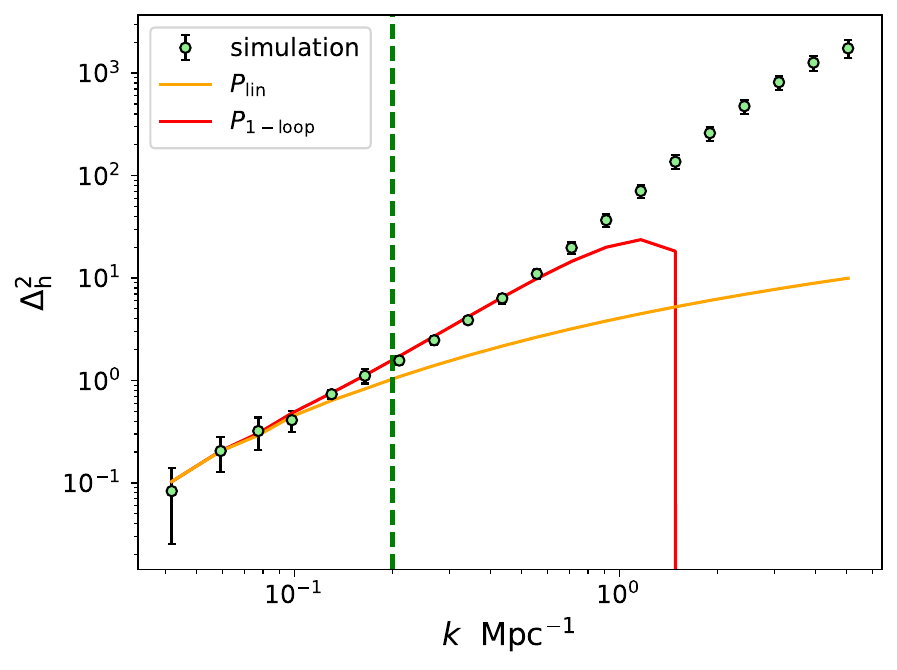} 
        \hspace{0.1cm}
        \includegraphics[width=0.325\textwidth]{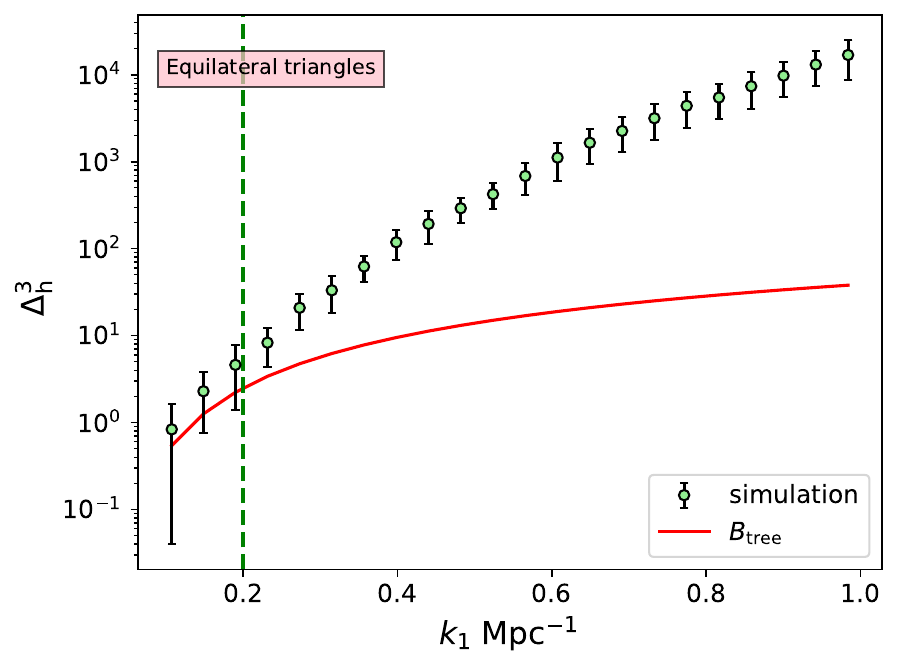}
        \hspace{0.1cm}
        \includegraphics[width=0.325\textwidth]{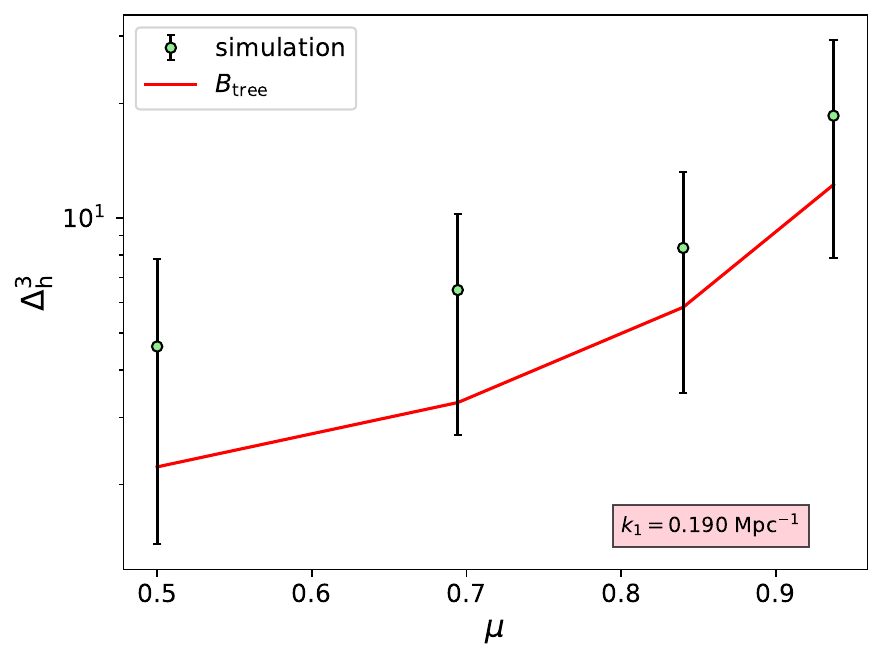}   
        
    \caption{Dimensionless Halo PS (left) and dimensionless Halo BS as a function of $k_1$ (middle) and $\mu$ (right). The green points with error bars represent the simulated PS and BS, and the red lines are the best-fit models. The orange line in the left panel is the linear PS scaled with $(b_1^{\rm eff})^2$. The green dashed lines are $k_{ul}$, and the points on the left of these lines have been used for the fitting.}
    \label{fig:halo_psbs}
\end{figure*}

\begin{figure}
\centering
	\includegraphics[width=\columnwidth]{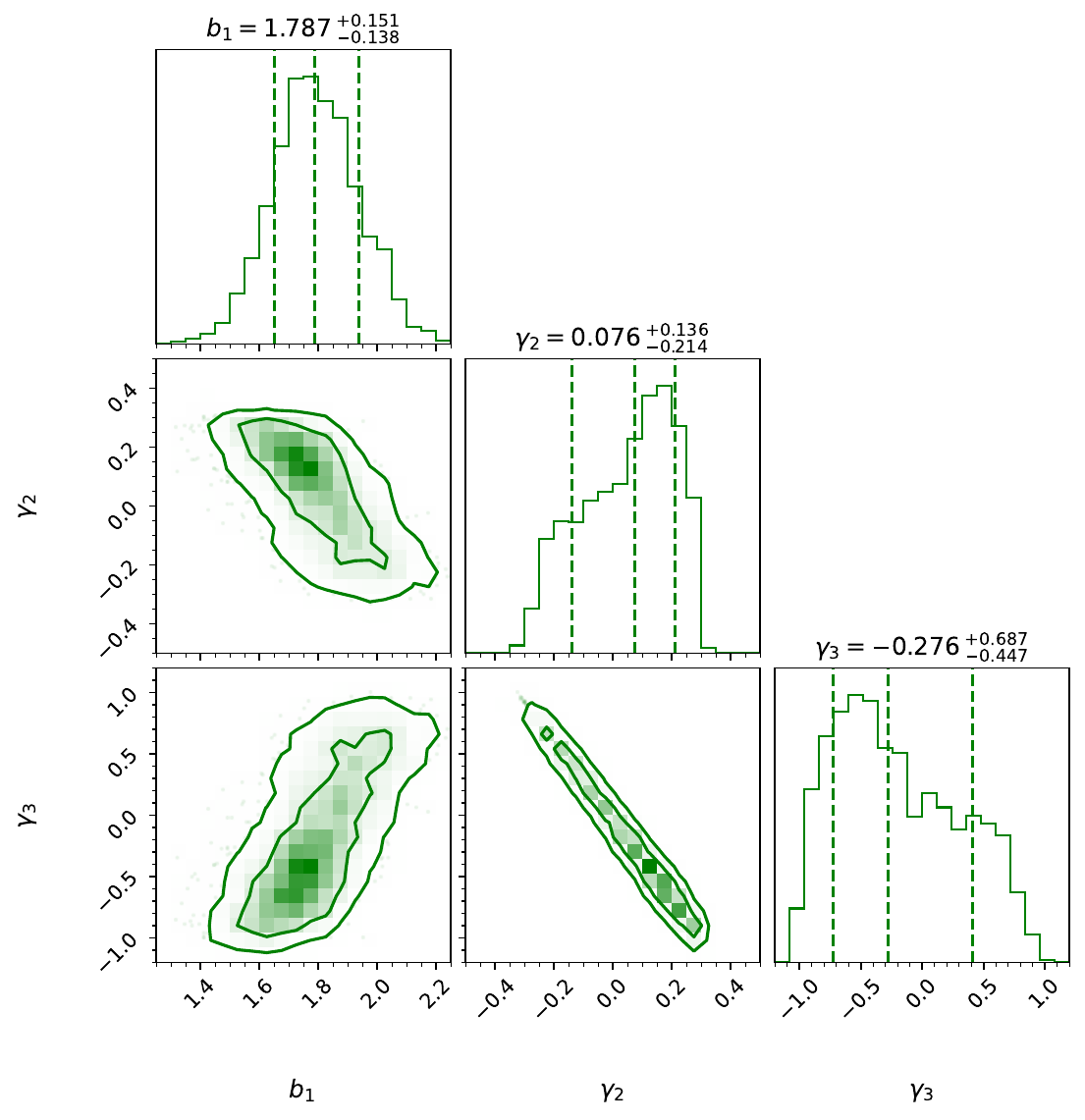}
    \caption{Posterior distribution for $b_1, \, \gamma_2, \, \gamma_3$ obtained using the simultaneous modeling of the Halo PS and BS using equations (\ref{eq:Ptr}) and (\ref{eq:Btr}).}
    \label{fig:halo_bias}
\end{figure}
We now consider the PS and BS of the simulated halo distribution at $z=1$. The halos 
are the same as those used to simulate the HI distribution in Section~\ref{sec:HI model}. 
These halos have a minimum mass $\sim 10^9 M_\odot$ set by the mass resolution of our simulations. Here, we have gridded the halo masses and converted this to a density distribution whose statistics we have analyzed. 
{Note that we have estimated the halo PS and BS using the halo mass density rather than the number density to account for the contributions of all halos in our simulation. The cumulative PS of the number density of halos with varying masses may be misleading, due to the fact that the halo PS changes with halo mass.}
Fig. \ref{fig:halo_psbs} shows $\Delta^2_{\rm h}$ the dimensionless halo PS (left panel) and $\Delta^3_{\rm h}$ the dimensionless halo BS (middle and right panels). The green points with error bars are respectively the mean and the $1 \sigma$ standard deviation, evaluated using $10$ independent realizations of the halo catalogue.   

In the left panel, we see that $\Delta^2_{\rm h} \approx 0.1$ at the smallest bin $k = 0.04 \, {\rm Mpc}^{-1}$,  and it increases to $\Delta^2_{\rm h}  \sim 10^3$ at $k \approx 3 \, {\rm Mpc}^{-1}$. The middle panel shows $\Delta^3_{\rm h}$ for equilateral triangles, with $k_1$ ranging from $k_1 = 0.11 \, {\rm Mpc}^{-1}$ to $ 0.98 \, {\rm Mpc}^{-1}$. We see that $\Delta^3_{\rm h}$ is positive and increases from $\sim 1$  to $\sim 10^4$ in this $k$ range. The right panel shows the $\Delta^3_{\rm h}$ for the L-isosceles triangles,  with two largest sides having the value $k_1 = k_2 = 0.190 \, {\rm Mpc}^{-1}$. We see that $\Delta^3_{\rm h}$  increases as the triangle shape changes from equilateral to squeezed.

\begin{figure*}
\centering
	\includegraphics[width=0.325\textwidth]{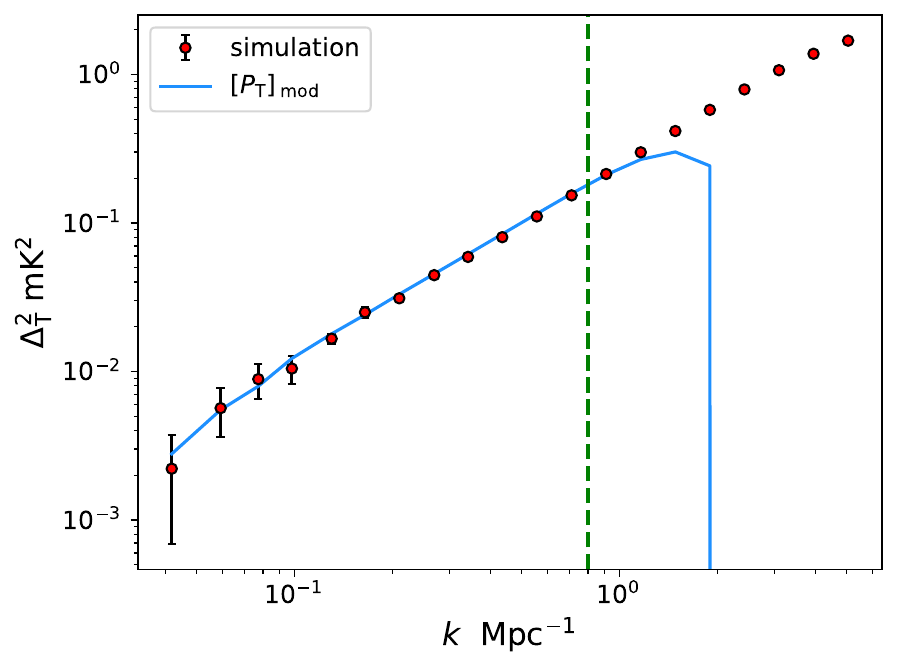} 
        \hspace{0.1cm}
        \includegraphics[width=0.325\textwidth]{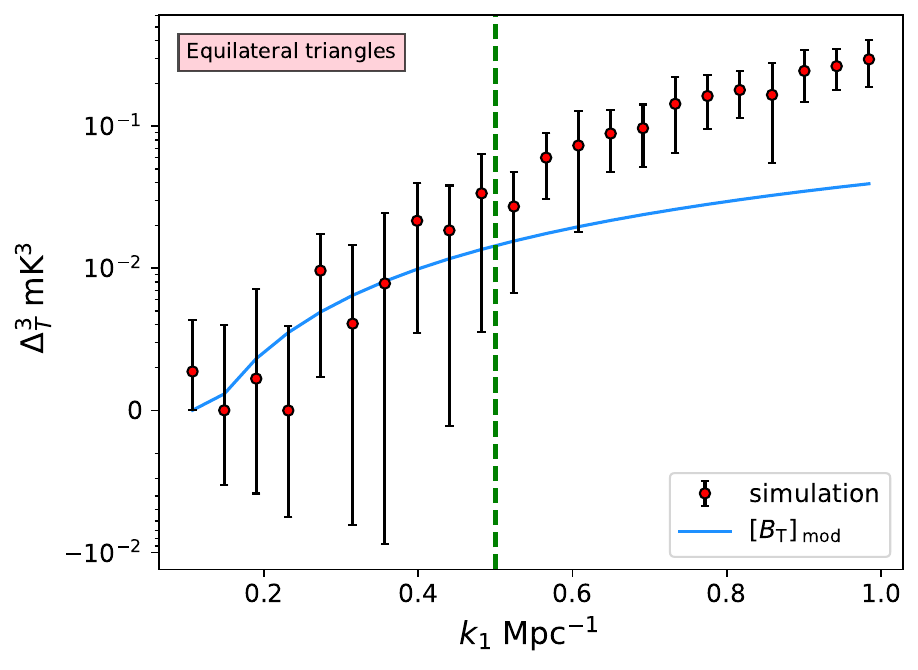}
        \hspace{0.1cm}
        \includegraphics[width=0.325\textwidth]{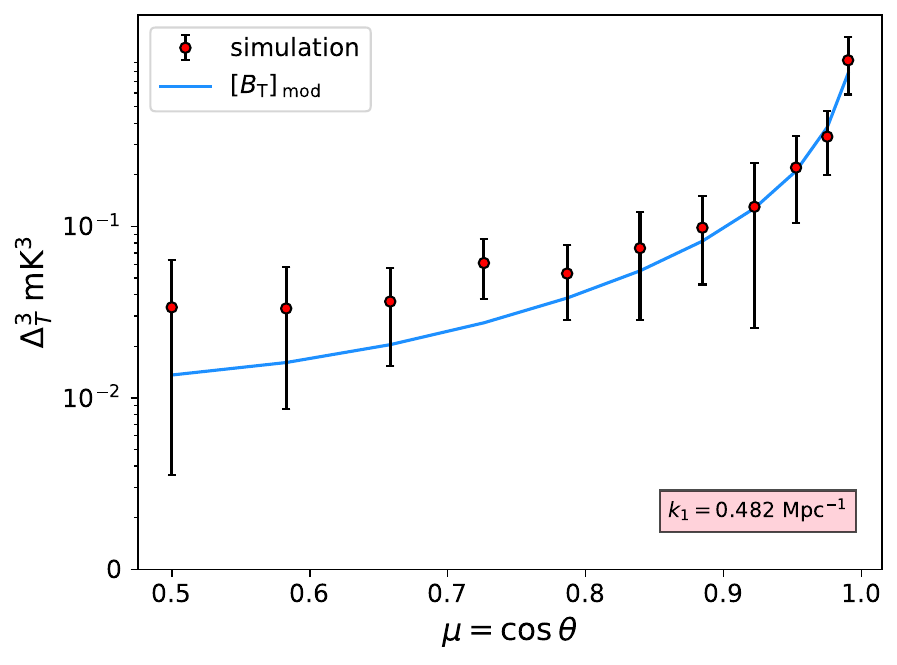}   
        
    \caption{Mean-squared (left) and mean-cubed (middle and right) brightness temperature fluctuations. Red points with error bars are the simulated values. The blue solid lines indicate the best-fit models (eq. \ref{eq:psbs_T}) and the green dashed lines indicate the extent of the fits.}
    \label{fig:h1_psbs}
\end{figure*}
\begin{figure}
\centering
	\includegraphics[width=\columnwidth]{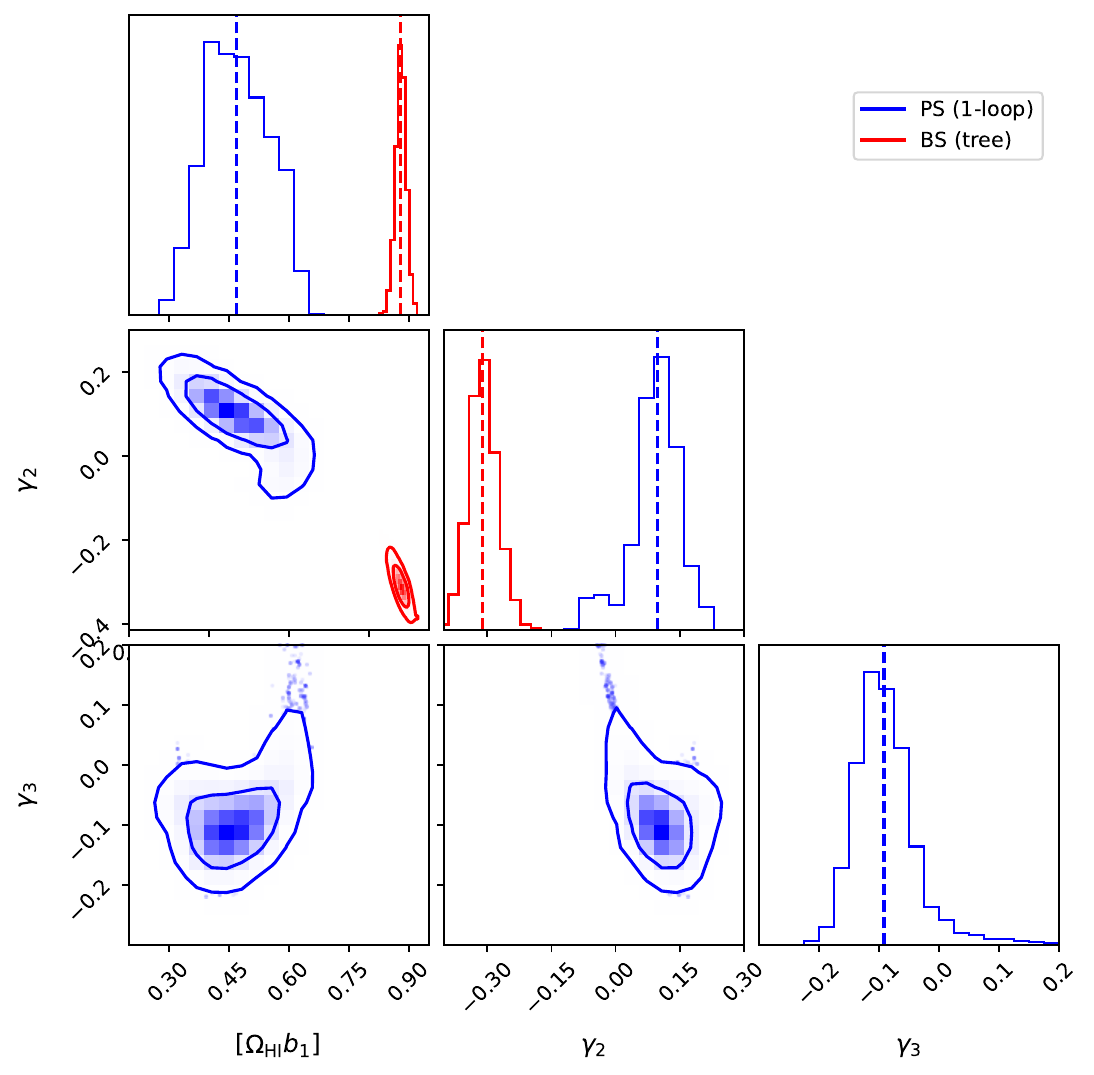}
    \caption{Posterior distributions for parameters $[\Omega_{\rm HI} b_1]$, $\gamma_2$, and $\gamma_3$ estimated by separately modeling the 21-cm PS and BS using the 1-loop PS (blue) and the tree-level BS (red) models. The dashed lines show the best-fit values (medians of the marginalized distributions).}
    \label{fig:HI_bias}
\end{figure}

We simultaneously model the PS and BS of the halo distribution using equations (\ref{eq:Ptr}) and (\ref{eq:Btr}). The modeling is restricted to large scales by limiting the k-range within  $k \leq k_{ul}$. By trial and error, we find that these equations are in good agreement with the simulations up to $k_{ul} = 0.2 \, {\rm Mpc}^{-1}$,  and they fail at larger $k$. We expect the linear bias to be greater than unity, since the halos are believed to reside at the peaks of the matter density field \citep{sheth1999large}. We use an MCMC analysis, similar to the one described in Section~\ref{subsec:bias mcmc}, to estimate the bias parameters $(b_1, \, \gamma_2, \, \gamma_3)$ for the halo distribution. We sample a Gaussian likelihood with flat priors for 2000 steps of 16 walkers and discard the first $10 \%$ samples as burn-in. 
\begin{table}
	\centering
	\caption{Prior ranges, best-fit values and $1 \sigma$ errors for the estimated $b_1$, $\gamma_2$ and $\gamma_3$ obtained using the Halo PS and BS.}
	\label{tab:bestfits_halo}
	\begin{tabular}{lcc} 
        \hline
		Parameter & Prior & Best-fit\\
		\hline
		  $ b_1$  & $[0,10]$ & $1.787 \, ^{+ 0.151}_{-0.138} $\\\\
     	$\gamma_2$ & $[-5,5]$ & $0.076 \, ^{+ 0.136}_{-0.214} $\\\\
        $\gamma_3$ & $[-5,5]$ & $-0.276 \, ^{+ 0.687}_{-0.447} $ \\
        \hline
	\end{tabular}
\end{table}
The posterior distribution, shown in Fig. \ref{fig:halo_bias}, exhibits tight constraints on all three parameters. The 2D contours show $68$ per cent and $95$ per cent confidence intervals, and the best-fit values of the bias parameters are chosen to be medians of the respective 1D marginalized distributions. The best-fit values and the $68$ per cent confidence intervals are quoted in Table \ref{tab:bestfits_halo}.We note that for $\gamma_2$ and $\gamma_3$,  the  $68$ per cent confidence intervals encompasses the value zero. 
The red lines in the left panel of Fig. \ref{fig:halo_psbs} indicate the best-fit 1-loop PS (eq \ref{eq:Ptr}) and best-fit tree-level BS (eq \ref{eq:Btr}) in the middle and right panels. We see that these models provide a good fit for $k < k_{ul}$, where the value of $k_{ul}$ is indicated by the green dashed lines.

Considering eq.~(\ref{eq:Ptr}), we expect the mode-coupling terms to approach zero in the limit $k \rightarrow 0$ where it reduces to 
\begin{equation}
    \label{eq:Ptr_l}
    P_{\rm 1-loop}  \,  (k) =
    b_1^2  \,  \left[ 1 + \left( \frac{68}{21} \gamma_2 + \gamma_3 \right) \sigma^2 \right] P(k) \,.
\end{equation}    
This allows us to define an effective "linear" bias 
\begin{equation}
    \label{eq:b1eff}
    b_1^{\rm eff} = b_1 \sqrt{ 1 + \left( \frac{68}{21} \gamma_2 + \gamma_3 \right) \sigma^2 } \, ,
\end{equation}
This effective linear bias exactly matches the definition used by \citet{nishizawa2013perturbation}, although they have arrived at it using slightly different considerations. Considering the posterior distributions of $b_1,\gamma_2$ and $\gamma_3$, we obtain the best-fit value $b_1^{\rm eff} = 1.393 \, ^{+ 0.089}_{-0.123}$.  The orange line in the left panel of Fig. \ref{fig:halo_psbs} shows  $(b_1^{\rm eff})^2 \, P(k) \, $. We see that the orange curve matches the simulation at $k \le  0.1 \, {\rm Mpc}^{-1}$, and deviates at larger $k$.  Furthermore, we see that the full 1-loop predictions (eq \ref{eq:Ptr}) match the simulations up to $k \approx 0.2 \, {\rm Mpc}^{-1}$.

We further validate our analysis by comparing our estimate of the effective linear bias  with the linear bias of the simulated halos, which  can also be analytically calculated by taking the cumulative bias contribution of all halos of different masses \citep{valageas2011large},
{
\begin{equation}
    \label{eq:cu_b1}
    \langle b_1 \rangle = \frac{\int_{M_{\rm min}}^\infty dM\, M \, n(M) \, b_{\rm h}(M)}{\int_{M_{\rm min}}^\infty dM \, M \, n(M)} \, .
\end{equation}
}
Here, $n(M)$ is the halo mass function {defined as the number density of halos per unit mass} and $b_{\rm h} (M)$ is the linear bias for halos of mass $M$. Both these quantities are taken to be of the Sheth-Torman (ST) form \citep{sheth1999large, sheth2001ellipsoidal} at $z=1$. We have used $M_{\rm min} = 10^9 M_\odot$,  which includes the contribution of all the halos in our simulation. We find that including an upper limit on the halo mass does not affect the final result, and we have considered this to be infinity. We find that the predicted cumulative bias has a value $\langle b_1 \rangle = 1.47$, which lies within the $68$ per cent confidence of our estimated $b_1^{\rm eff}$. 

We now attempt to jointly model the 21-cm PS and BS (described in Sections~\ref{sec:HI model} and \ref{sec:Bias est}) using equations (\ref{eq:Ptr}) and (\ref{eq:Btr}). For this, we introduce appropriate scaling factors to define 
\begin{equation}
    \label{eq:psbs_T}
    \begin{aligned}
    & [P_{\rm T}]_{\rm mod} =  \left(\frac{4 \bar{T}}{3 \Omega_b}\right)^2 \Omega_{\rm HI}^2 \, P_{\rm 1-loop}\\ & {\rm and} \\
    & [B_{\rm T}]_{\rm mod} = \left(\frac{4 \bar{T}}{3 \Omega_b}\right)^3 \Omega_{\rm HI}^3 \,  B_{\rm tree}
    \end{aligned}  
\end{equation}
where the $[B_\text{T}]_{\rm mod}$ is the same as in eq. (\ref{eq:bs}). The free parameters of the 21-cm PS and BS models are $[\Omega_{\rm HI} b_1]$, $\gamma_2$, and $\gamma_3$. The simulated 21-cm PS and BS are described in the Section~\ref{subsec:PS BS}. We start by separately modeling the PS and BS using the above equations and comparing the estimated parameter values. Note that the BS can be used to only model $[\Omega_{\rm HI} b_1]$ and $\gamma_2$ and has no contribution from $\gamma_3$. Using separate $\chi^2$ minimization (eq. \ref{eq:chisq}), we find that the PS can be modeled for $k \le 0.8 \, {\rm Mpc}^{-1}$ and BS for $k \le 0.5 \, {\rm Mpc}^{-1}$. Note that here we have used only the diagonal elements of the covariance matrix $\mathbf{C}$ as we do not obtain reliable estimates of the off-diagonal elements from the limited number of realizations of the simulated signal. 
We now perform an MCMC analysis similar to the one described for the halo distribution, sampling a Gaussian likelihood for 1000 steps with 16 random walkers and discarding the first $10$ per cent as burn-in. The best-fit values (medians) and their $68$ per cent confidence intervals are indicated in Table \ref{tab:bestfits_h1psbs}. 

The left panel Fig. \ref{fig:h1_psbs} shows $\Delta_{\rm T}^2$ the simulated mean-squared brightness temperature fluctuations (red points, same as fig. \ref{fig:ps}). The middle and right panels show $\Delta_{\rm T}^3$ the mean-cubed brightness temperature fluctuations. The middle panel shows $\Delta_{\rm T}^3$ for the equilateral triangles (same as the top panel of fig. \ref{fig:bs}) and right panel shows $\Delta_{\rm T}^3$ for L-isosceles triangles with two largest sides $k_1 = k_2 = 0.482 \, {\rm Mpc}^{-1}$. Similar to the bottom panel of fig. \ref{fig:bs}, $\Delta_{\rm T}^3$ increases as the triangle shape changes from equilateral to squeezed. The best fit models (eq. \ref{eq:psbs_T}) are indicated by the blue solid lines. We see that these models are a good fit to the 21-cm PS for $k < 0.8 \, {\rm Mpc}^{-1}$ and BS for $k < 0.5 \, {\rm Mpc}^{-1}$ (indicated by green dashed lines).
The posterior distributions for the PS (blue) and BS (red) are shown in Fig. \ref{fig:HI_bias}.
\begin{table}
	\centering
	\caption{Prior ranges, best-fit values and $1 \sigma$ errors for the estimated $b_1$, $\gamma_2$ and $\gamma_3$ using separate analysis of the 21-cm PS (1-loop) and BS (tree-level).}
	\label{tab:bestfits_h1psbs}
	\begin{tabular}{cccc} 
        \hline
		Parameter & Prior & PS (1-loop) & BS (tree)\\
		\hline
		  $ [\Omega_{\rm HI} b_1] (\times 10^{-3})$  & $[0,1]$ & $0.47 \, ^{+ 0.09}_{-0.08} $ & $0.88 \, ^{+ 0.01}_{-0.01} $\\\\
     	$\gamma_2$ & $[-5,5]$ & $0.10 \, ^{+ 0.05}_{-0.05} $ & $-0.31 \, ^{+ 0.04}_{-0.04} $\\\\
        $\gamma_3$ & $[-5,5]$ & $-0.09 \, ^{+ 0.06}_{-0.04} $ & $ - $\\
        \hline
	\end{tabular}
\end{table}
Comparing the two distributions, we see that they are centered at very different positions in the parameter space. This indicates that the PS and BS prefer very different values of $[\Omega_{\rm HI} b_1]$ and $\gamma_2$ for the 1-loop PS and tree-level BS models. We find that due to competing bias preferences of the 1-loop PS and tree-level BS for the 21-cm distribution, it is not possible to simultaneously model the two using equations \ref{eq:psbs_T}. The MCMC chain does not converge for such a situation, even for small $k < 0.25 \, {\rm Mpc}^{-1}$. 

From the above analysis, we find that the combination of 1-loop PS and tree-level BS can simultaneously model the matter and halo distributions. The 21-cm PS and BS can also be modelled separately using equation \ref{eq:psbs_T}, but this yields very different values for the bias parameters. Therefore, it is not feasible to simultaneously model the 21-cm PS and BS using this set of equations. We might be able to solve this compatibility issue by using the 1-loop BS \citep{scoccimarro1997cosmological, scoccimarro1998nonlinear}. However, the 1-loop matter BS contains SPT kernels up to the fourth order. We then have to include a quartic bias 
in eq.~(\ref{eq:bias_3}), and also include the two-loop correction to the PS in eq.~(\ref{eq:Ptr}).  This will lead to a huge number of terms in the expressions for the PS and BS, which might be hard to model with the limited data points. The tracer PS and BS can also be modeled using the Effective Field Theory (EFTofLSS) approach (\cite{mcdonald2009clustering, baumann2012cosmological, senatore2015ir, angulo2015statistics, akitsu2025cosmology}, see \cite{ivanov2024effective} for a review). This approach has been extensively used to model the observed galaxy distribution \citep{gil2015power, colas2020efficient, ivanov2020cosmological}. It includes contributions from tidal effects and local velocity terms in the tracer overdensity, effectively rendering the contribution from the underlying matter density non-local and increasing the number of free bias parameters to $> 5$. This is again difficult to estimate using a limited number of points on the relevant large scales. Furthermore, it will be difficult to relate these parameters to the parameters of the HIHM relation. 

We therefore conclude that to simultaneously model the 21-cm PS and BS, it is best to combine the linear model for PS and the tree-level BS (equations \ref{eq:ps} and \ref{eq:bs}) as shown in Section~\ref{subsec:bias-model}. 


\section{Joint PS and BS correlation matrix}
\label{ap:correlation}

\begin{figure}
\centering
	\includegraphics[width=\columnwidth]{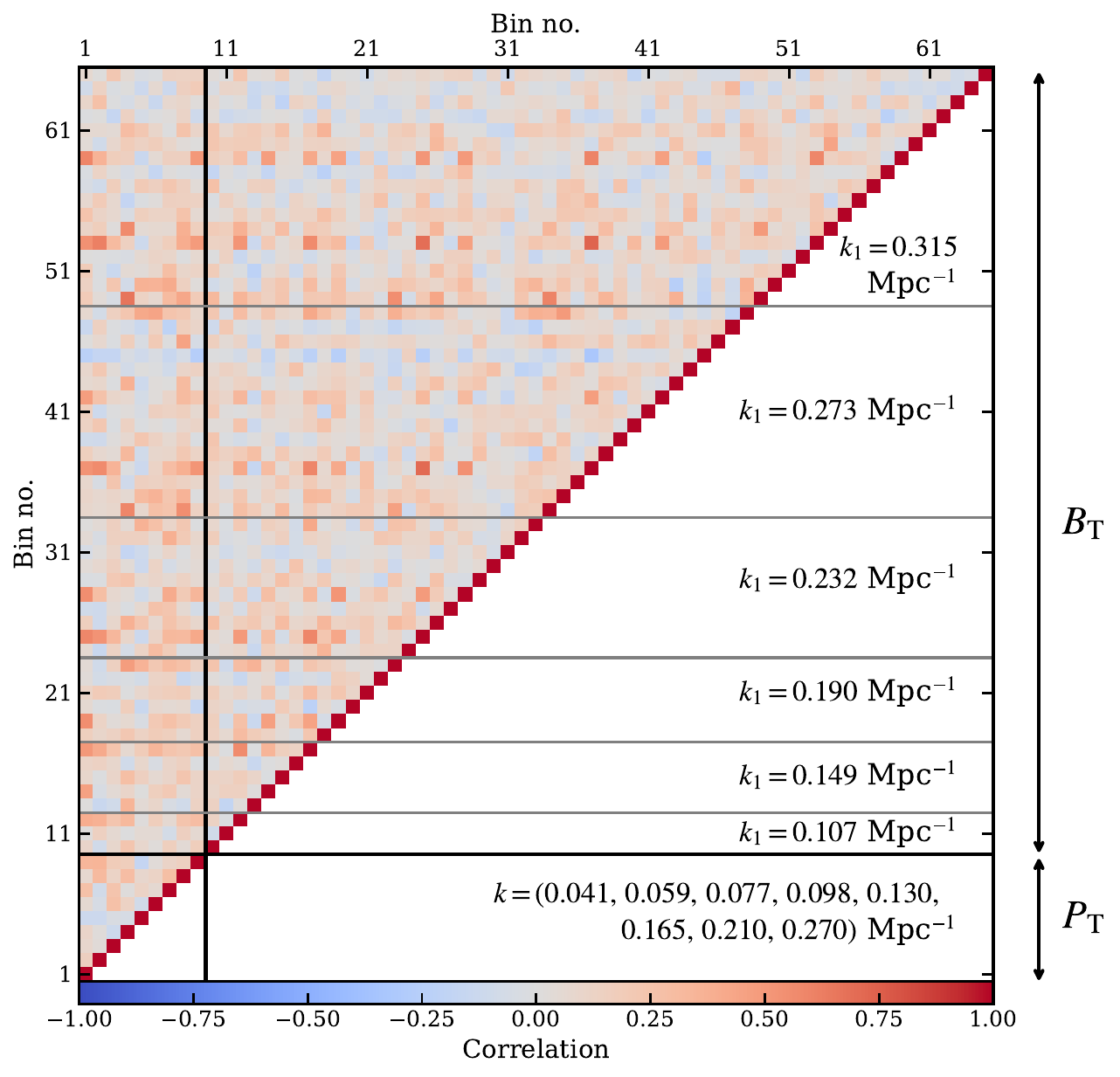}
    \caption{The upper triangle of the symmetric $65 \times 65$ dimensionless correlation matrix for the 21-cm PS and BS obtained using 100 statistically independent realizations of the simulated HI distribution. Bin numbers $1-8$ correspond to $P_\text{T}(k)$, the PS,  where the corresponding $k$ values are shown in the figure. Bin numbers $9-65$ correspond to $B_\text{T}(k_1,\mu,t)$, the BS. The BS bins are divided into $6 \, k_1$ stacks separated by gray horizontal lines. Each stack corresponds to a fixed $k_1$, which is indicated in the figure. The values of $(\mu,t)$ change across the bins within a stack, covering triangles of different shapes. Within a stack,  the largest bin number corresponds to equilateral triangles. The triangle shape changes to squeezed and ends with stretched  as the bin number decreases.}
\end{figure}

In this section, we present a visualisation of the dimensionless correlation matrix $r_{ij}=C_{ij} \, /\sqrt{C_{ii} C_{jj}}$ for the covariance matrix $C_{ij} \equiv \mathbf{C}$.

\end{document}